%% file: paper_V2.tex
\documentclass[fleqn,usenatbib]{mnras}

\usepackage{newtxtext,newtxmath}

\usepackage[T1]{fontenc}

\DeclareRobustCommand{\VAN}[3]{#2}
\let\VANthebibliography\thebibliography
\def\thebibliography{\DeclareRobustCommand{\VAN}[3]{##3}\VANthebibliography}

\usepackage{graphicx}	
\usepackage{amsmath}	
\usepackage{xfrac}
\usepackage{xspace}
\usepackage{caption}
\usepackage{subcaption}
\usepackage{soul}
\setstcolor{red}

\newcommand{\Abacus}{\textsc{Abacus}\xspace}
\newcommand{\AbacusSummit}{\textsc{AbacusSummit}\xspace}

\title[Accuracy of pairwise velocities]{Constraining accuracy of the pairwise velocities in $N$-body simulations using scale-free models}

\author[S. Maleubre et al.]{
Sara Maleubre,$^{1,2}$\thanks{E-mail: sara.maleubre@lpnhe.in2p3.fr}
Daniel J.\ Eisenstein,$^{3}$
Lehman H.\ Garrison,$^{4,5}$
and Michael Joyce$^{1}$ \\
$^{1}$ Laboratoire de Physique Nucléaire et de Hautes Énergies, UPMC IN2P3 CNRS UMR 7585, \\ Sorbonne Université, 4, place Jussieu, 75252 Paris Cedex 05, France \\
$^{2}$ Max Planck Institute for Extraterrestrial Physics, Giessenbachstrasse 1, 85748 Garching, Germany \\
$^{3}$ Center for Astrophysics $|$ Harvard $\&$ Smithsonian, 60 Garden St, Cambridge, MA 02138  \\
$^{4}$ Center for Computational Astrophysics, Flatiron Institute, 162 Fifth Ave., New York, NY 10010 \\
$^{5}$ Scientific Computing Core, Flatiron Institute, 162 Fifth Ave., New York, NY 10010 \\
}

\date{Accepted XXX. Received YYY; in original form ZZZ}

\pubyear{2022}

\begin{document}
\label{firstpage}
\pagerange{\pageref{firstpage}--\pageref{lastpage}}
\maketitle

\begin{abstract}
{
We present a continuation of an analysis that aims to quantify resolution of \emph{N}-body simulations by exploiting large (up to $N=4096^3$) simulations of scale-free cosmologies run using \Abacus. Here we focus on radial pairwise velocities of the matter field, both by direct estimation and through the cumulative-2PCF (using the pair conservation equation). We find that convergence at the 
$1\%$ level of the mean relative pairwise velocity can be demonstrated over a range of scales, 
evolving from a few times the grid spacing at early times to slightly below this scale at late times. 
We show the analysis of two different box sizes as well as from averaging results from the smaller boxes, and compare the power of the two aforementioned estimators in constraining accuracy at each scale.
Down to scales of order of the smoothing parameter,
convergence is obtained at $\sim5\%$ precision, and shows 
a behaviour indicating asymptotic stable clustering. We also infer for LCDM simulations conservative estimates on the evolution of the lower cut-off to resolution 
(at $1\%$ and $5\%$ precision) as a function of redshift.
}
\end{abstract}




\section{Introduction}\label{sec:Intro}
\input{Sections/1_Introduction.tex}

\section{Scale-Free Simulations and PairWise Velocity}\label{sec:PWandSF}
\input{Sections/2_PWvel_and_SFsims.tex}

\section{Numerical simulations}\label{sec:NumSim}
\input{Sections/3_NumericalSimulations.tex}

\section{Results}\label{sec:Results}
\input{Sections/4_Results.tex}

\section{Conclusions}\label{sec:conclusions}
\input{Sections/6_Conclusions.tex}



\section*{Acknowledgements}

S.M. thanks the Institute for Theory and Computation (ITC) and the Flatiron Institute for hosting her in early 2022, and acknowledges the Fondation CFM pour la Recherche and the German Academic Exchange Service (DAAD) for financial support. S.M. and M.J. thank Pauline Zarrouk for useful discussions.

D.J.E. is supported by U.S.\ Department of Energy grant, now DE-SC0007881, NASA ROSES grant 12-EUCLID12-0004, and as a Simons Foundation Investigator.

This research used resources of the Oak Ridge Leadership Computing Facility at the Oak Ridge National Laboratory, which is supported by the Office of Science of the U.S. Department of Energy under Contract No. DE-AC05-00OR22725. The \AbacusSummit simulations have been supported by OLCF projects AST135 and AST145, the latter through the U.S.\ Department of Energy ALCC program.

\section*{Data Availability}

Data access for the simulations part of \AbacusSummit is available through OLCF’s Constellation
portal. The persistent DOI describing the data release is
\href{https://doi.ccs.ornl.gov/ui/doi/355}{10.13139/OLCF/1811689}. Instructions for accessing the
data are given at \href{https://abacussummit.readthedocs.io/en/latest/data-access.html}{https://abacussummit.readthedocs.io/en/latest/data-access.html}.

Data corresponding to the smaller simulations as well as the derived data generated in this research will be shared on reasonable request to the corresponding author.



\bibliographystyle{mnras}
\bibliography{Bibliography} 




\appendix

\section{Mapping from EdS to LCDM-like cosmologies}\label{app:copyPS}
\input{Sections/Appendix.tex}



\bsp	
\label{lastpage}
\end{document}

%% file: Sections/1_Introduction.tex
Observational tests such as Type Ia supernovae \citep{Perlmutter1997, Riess1998}, large-scale structure analysis from Baryon Acoustic Oscillations \citep[BAO,][]{Eisenstein2005, Cole2005} and the temperature anisotropies of the cosmic microwave background \citep[CMB,][]{Jaffe2001, Pryke2002, planck_2013} provide compelling evidence that the Universe is in an accelerated expansion. To explain this within the framework of General Relativity requires a new type of ``dark'' energy  that accounts for about $70\%$ of the total, and whose nature is still unknown. In the current standard model of cosmology (LCDM), this energy component is in the form of a cosmological constant.
Alternative theoretical approaches 
either add extra degrees of freedom to characterize the energy content of the Universe or modify the Einstein-Hilbert action \citep[for a review on these models see][] {Clifton20121}. 

Ongoing and future surveys such as the Dark Energy Spectroscopic Survey (DESI) \citep{DESI} or the space-based mission \emph{Euclid} \citep{Euclid} will provide large scale structure maps of the Universe of unprecedented statistical precision, allowing astronomers to measure the expansion history of the Universe and the growth rate of cosmic structures in sufficient detail to potentially distinguish between the different possible aforementioned scenarios.

Indeed, one of the most valuable tests to discriminate between these multiple models observationally, and ultimately determine which can explain current data, consists in the study of the rate at which cosmic structures grow \citep[see e.g.][]{Perenon2019, Brando2021}, as different theories can predict quite different growth histories even for the same background evolution. A popular way of constraining this growth rate is by analysing the corrections to galaxy redshifts due to their peculiar velocities, which produces a modification of galaxy clustering, an effect called redshift-space distortions \citep[RSD,][]{Jackson1972, Kaiser1987}. Since peculiar velocities are caused by gravitational pull, we can trace a relation between the velocity field and the mass density field and thus estimate the rate at which structures grow. Countless efforts have been made into modelling these velocities from low-order statistics of the density field (in particular from the 2PCF) \citep[e.g.][]{Mo1997, Juszkiewicz1999, Sheth2001a,Sheth2001b}, as well as higher order multiples \citep{Scoccimarro2004,Shirasaki2021,Carolina2020}

In order to exploit this information, it is essential to calculate accurate theoretical predictions for the large-scale structure of the Universe. Below scales where the perturbative approaches break down, such calculations rely entirely on cosmological simulations performed using the $N$-body method. This approach approximates the continuous phase-space distribution of dark matter by that of a sparse finite sample of particles, and evolves them in a finite box with periodic boundary conditions. In this context, an important question is the accuracy and scale-range limitations of this method in attaining the physical limit.

The assessment of the accuracy to which results converge to values independent of the 
numerical parameters (time stepping, force accuracy parameters) introduced in the 
resolution of the $N$-body system is straightforward. In this respect, extensive 
code comparisons \citep{Heitmann2008,Schneider2016,Garrison2019,Grove2022} give 
considerable added confidence in the precision of results for different statistics.
Such comparisons do not address, however, the question of the accuracy with which 
these simulations represent the physical limit. While dependence on box size can be assessed by 
direct extrapolation studies \citep[see e.g.][]{Knabenhans2019}, assessing the accuracy 
limitations imposed {\it at small scales} due to the discretization 
of the matter field is much more complex. The reason is that there are, at least, two relevant unphysical parameters, the mean interparticle spacing (denoted $\Lambda$ here) and the gravitational force smoothing 
(denoted $\epsilon$), and numerical extrapolation to the continuum physical limit, corresponding to $\epsilon/\Lambda \rightarrow 0$, is in practice unattainable. Precise quantitative conclusions 
regarding it have remained elusive and sometimes controversial  \cite[see][for a discussion and some references]{Joyce2021}. 

Previous studies using $N$-body simulations have already used the information contain in the dark matter and halo pairwise velocity field to study plausible deviations from the standard model \citep{Hellwing2014,Gronke2015,Bibiano2017,Hellwing2017,Dupuy2019,Valogiannis2020}. Such conclusions ultimately rely on the ability of the $N$-body method to accurately predict and compute the desired statistic, and that of the chosen halo finder retrieving halo properties accurately. But halos are not uniquely defined entities, and their 
properties  depend strongly 
on the  algorithm adopted for their extraction. We will explore this topic in an accompanying paper \citep{Maleubre2023}.

In this article, we use the techniques introduced in \citet{Joyce2021} and developed and 
applied also in \citet{Leroy2021,Garrison2021,Garrison2021c} and \citet{Maleubre2022} to derive resolution limits  
arising from particle discretization for different statistics by analysing deviations from self-similarity in scale-free cosmological models. Here, we employ these methods
to assess and quantify the limits arising from discretization on the precision at which the radial component of the pairwise velocity of the full dark matter field can be retrieved from $N$-body simulations.  

This article is structured as follows. The first part of \autoref{sec:PWandSF} describes what scale-free cosmologies are and how their self-similar evolution can be used to determine the accuracy at which different statistics can be measured in $N$-body simulations. Next, we recall the expressions for the radial component of the pairwise velocity and the pair conservation equation, as well as give the equation for the latter in the context of scale-free cosmologies. \autoref{sec:NumSim} contains a summary of the simulations used, as well as a brief description of \Abacus, the $N$-body code used for their computation. It also contains a description of the method used to estimate convergence of the different statistics in the dark matter field. In \autoref{sec:Results} we present and analyse our results, as well as infer resolution limits to non-scale-free cosmologies. Finally, we summarize our main findings in \autoref{sec:conclusions}.

%% file: Sections/2_PWvel_and_SFsims.tex
\subsection{Scale-free simulations and Self Similarity}\label{sec:SSandSS}
Scale-free cosmologies have an Einstein-de Sitter, EdS, ($\Omega_{M}=1$) background and a power-law power spectrum ($P_{k}\propto k^{n}$) of initial perturbations, which are thus characterized by just one length scale, the scale of non-linearity. This can be defined by
\begin{equation}\label{eq:RNL}
    \sigma^2_{\rm lin}(R_{\rm NL},a)=1 
\end{equation}
where $\sigma^2_{\rm lin}$ is the variance of normalized linear mass fluctuations in a sphere. Its temporal evolution can be calculated from linear perturbation theory as  
\begin{equation}\label{eq:RNL-scaling}
    R_{\rm NL}\propto a^{\frac{2}{3+n}}
\end{equation}

One can infer that, if the evolution of gravitational clustering is independent of any other length scale (notably ultraviolet or infrared cut-offs to the assumed power-law fluctuations), it must be {\it self-similar}, i.e., the temporal evolution of the statistics describing clustering is given by a spatial rescaling following \autoref{eq:RNL-scaling}. More specifically, 
any dimensionless function $F(x_1,x_2,...;a)$ describing clustering (where the $x_i$ are the parameters on which the statistic depends)
will obey a relation of the form 
\begin{equation}\label{eq:SS-sgeneral}
    F(x_1,x_2,...;a)=F_0(x_i/X_{{\rm NL},i}(a))
\end{equation}
where $X_{{\rm NL},i}$ encodes the temporal dependence of the characteristic scale with the same dimensions as $x_i$ (as inferred from $R_{\rm NL}$). 

\begin{figure*}
  \begin{subfigure}{\linewidth}
    \includegraphics[width=\linewidth]{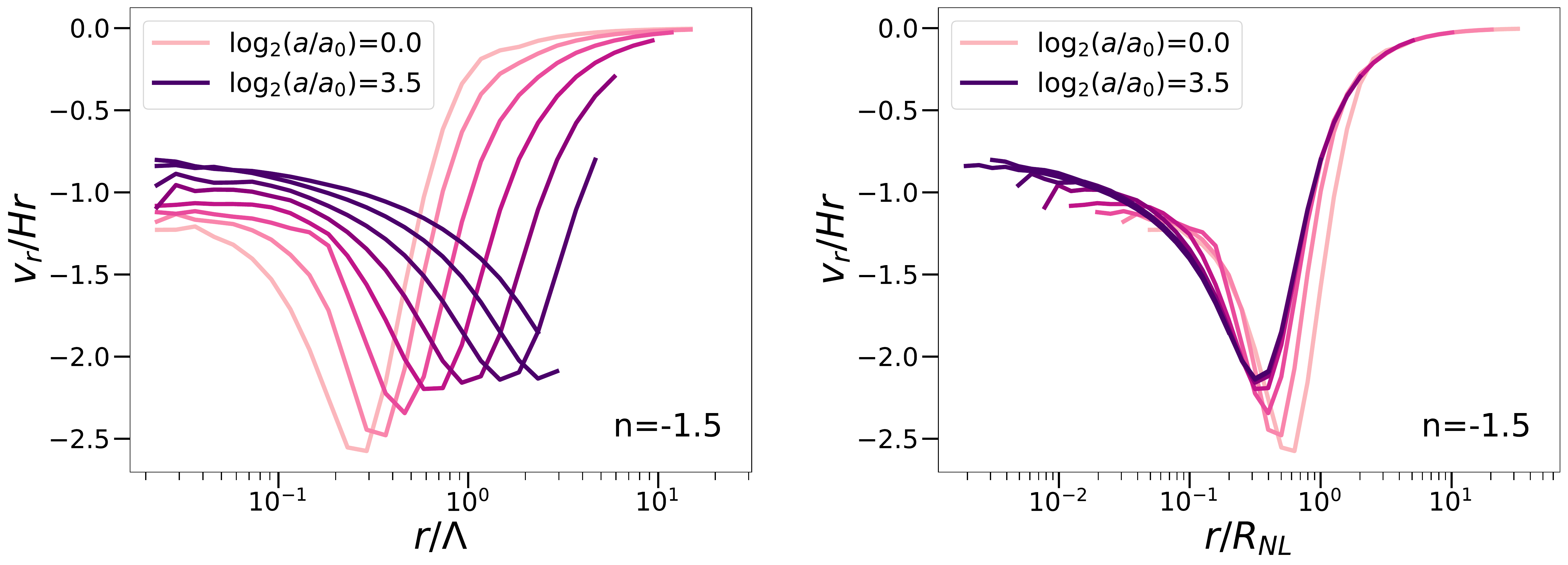}
  \end{subfigure}
  \begin{subfigure}{\linewidth}
    \includegraphics[width=\linewidth]{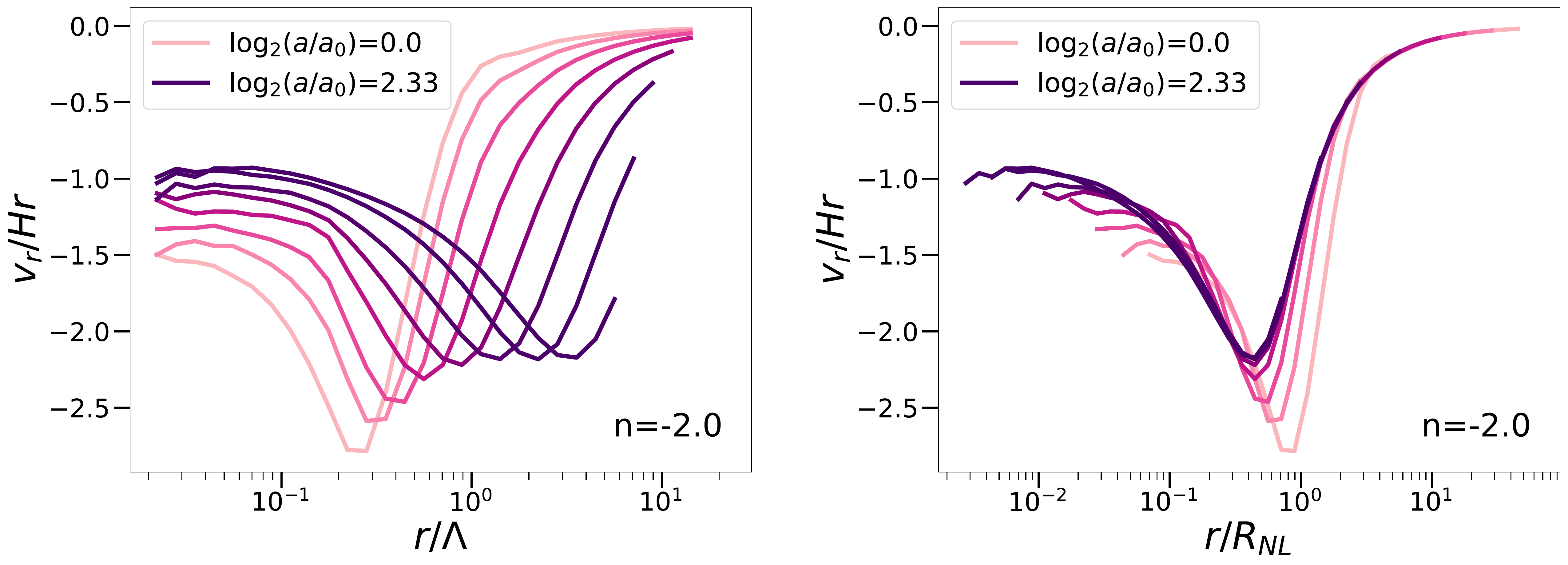}
  \end{subfigure}
  \begin{subfigure}{\linewidth}
    \includegraphics[width=\linewidth]{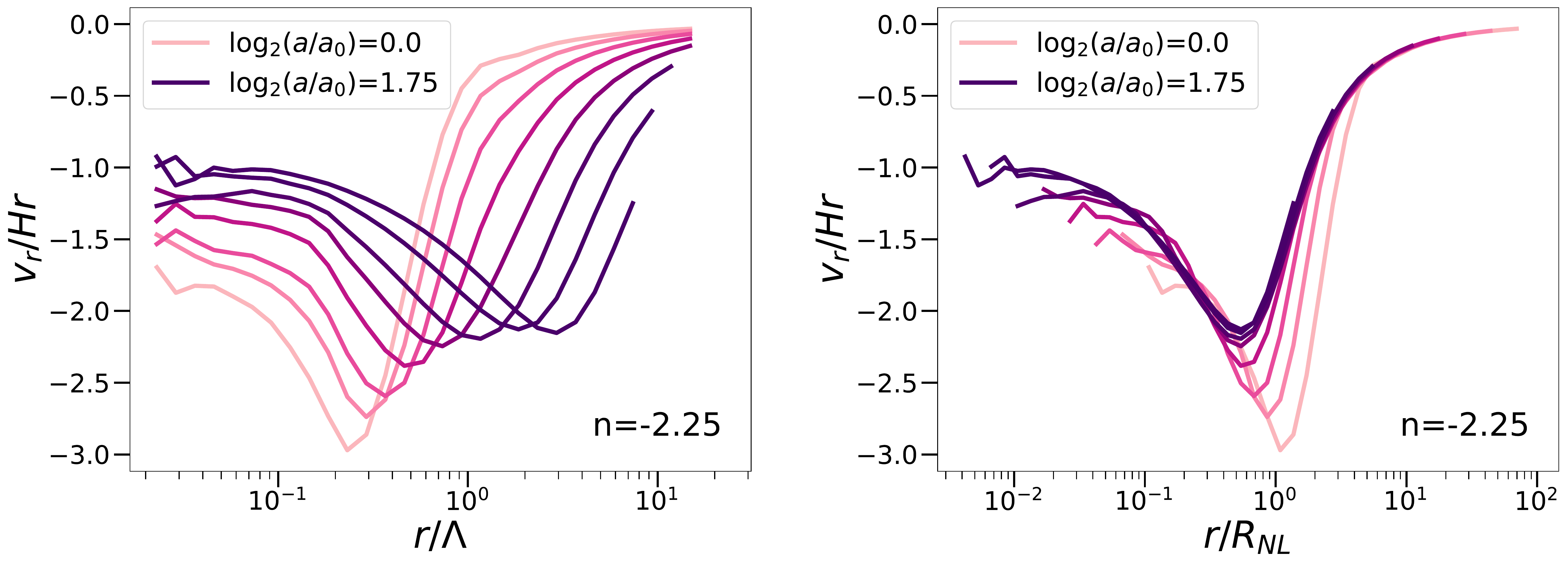}
  \end{subfigure}
  \caption{Directly estimated $v_{r}/Hr$ as a function of comoving separation (left column) and of the rescaled coordinate $r/R_{\rm NL}$ (right column), for simulations with spectral indices $n=-1.5$, $n=-2.0$ and $n=-2.25$ ($N=4096^3$ for the former exponent and average over four $N=1024^3$ for the last two). Self-similar evolution corresponds to a
  superposition of the curves in the rescaled plots. The times shown correspond to every fourth snapshot $S=0,4,8,...$ (where $S$ is as defined in \autoref{eq:Sdef}) over the total time-span of the simulations. }
  \label{fig:SS_SF}
\end{figure*}

Our interest in self-similarity is driven by the fact that it greatly simplifies the description of clustering: its time dependence is effectively trivial, and any statistic describing clustering is specified by the single time-independent function on the right-hand side of \autoref{eq:SS-sgeneral}.
As discussed in our previous papers, we can use this property to determine the range of scales that a simulation can reliably reproduce: any deviation from self-similarity arises necessarily from dependence on the \emph{unphysical scales} proper to the $N$-body simulations.

\subsection{Pairwise Velocity and pair-conservation equation}\label{sec:PW_PC}

In this study we focus on the radial component of the mean pairwise velocity defined by
\begin{equation}
    v^{r}_{12}=\left<(\mathbf{v}_1-\mathbf{v}_2)\cdot\frac{\mathbf{r}}{|\mathbf{r}|}\right>
\end{equation}
where the velocity difference $(\bf{v}_1-\bf{v}_2)$ of a pair of objects is projected on to their separation vector $\bf{r}$,
and $<\cdots >$ denotes the ensemble average.
It can be estimated in a finite simulation by directly averaging the pair velocity over all pairs. To do so, here we have coded an appropriate  
modification of the analysis tool \emph{Corrfunc} \citep{Corrfunc_extra,Corrfunc2020}. To facilitate our analysis based on self-similarity, 
we will always consider below the dimensionless ratio of $v^{r}_{12}$ to the Hubble flow ($Hr$), so that self-similarity has the 
simple expression in the form of \autoref{eq:SS-sgeneral}.

This paper focuses on the matter field, and the choice to study $v^{r}_{12}$ is motivated by the fact that, in this case, it can also be related to the two-point correlations of mass density via the so-called pair conservation equation. This relation was first derived by \citet{DandP1977} as a consequence of the BBGKY equations. In their statistical description, matter was approximated by a set of identical particles of mass \emph{m}, making their theoretical results directly applicable to those of $N$-body simulations. 
Starting from the continuity equation for the density contrast (zeroth moment of the Vlasov equation)
one obtains the pair conservation equation:
\begin{equation}
    \frac{\partial\xi_{12}}{\partial\tau}+\nabla_{12}\cdot\left[\mathbf{v}^{r}_{12}\left(1+\xi_{12}\right)\right]=0
\end{equation}
where $\tau$ is equal to the conformal time and $\xi_{12}$ is the standard reduced two-point density-density correlation function (2PCF) defined as the ensemble average at two different locations ($1+\xi_{12}=<(1+\delta(\mathbf{x}_1))(1+\delta(\mathbf{x}_2))>$).

This can be conveniently rewritten as \citep{Nityananda1993}:
\begin{equation}\label{eq:pc}
     \frac{v_r}{Hr}=-\frac{1}{3(1+\xi)}\frac{\partial\bar{\xi}}{\partial\ln a}
\end{equation}
where $\bar{\xi}=3x^{-3}\int_0^x\xi y^2dy$, the cumulative two-point correlation function (\emph{cumulative} 2PCF), is the average 2PCF interior to x where we have normalized the velocity to the Hubble flow ($Hr$). For economy, we have dropped the indices ${12}$ in the two-point quantities.  As \autoref{eq:pc} is exact,  it implies that we can estimate $v_r$ in a finite sample indirectly, using instead of the velocities themselves the direct estimators of the 2PCF, the \emph{cumulative} 2PCF and its derivative, combined in the appropriate way. This has been previously exploited in an early study of the pair velocity in scale-free models by \citet{Jain1997} focused on the question of whether clustering
become {\it stable} at small scales \citep{Peebles1974}, i.e. whether it tends to become stationary in 
physical coordinates, corresponding to $v_r=-Hr$.

In the context of scale-free models and their expected self similarity, it is convenient to rewrite \autoref{eq:pc} with the time derivative taken at a fixed value of the rescaled comoving separation  (i.e. at fixed $ r/R_{\rm NL}$ rather than fixed $r$)
\begin{equation}\label{eq:pw_pc}
    \left . \frac{v_r}{Hr}=-\frac{2}{3+n}\left(\frac{\bar{\xi}}{\xi}-1\right)\frac{\xi}{1+\xi}-\frac{1}{3\left(1+\xi\right)}\frac{\partial\bar{\xi}}{\partial\ln a}\right|_{r/R_{\rm NL}}\,.
\end{equation}
When the two-point density correlations (as described by $\xi$ and $\bar{\xi}$) are self-similar, the last term vanishes and we can infer that 
$v_r$ is also self-similar.
On the other hand, self-similarity of $\xi$ and $\bar{\xi}$ is not a requirement for that of $v_r$. We will pay careful attention to this point in our analysis below, and we will show that there is in fact a regime in our simulations in which $v_r$ approximates well self-similarity while the 2PCF does not.

%% file: Sections/3_NumericalSimulations.tex
\subsection{\Abacus code and simulation parameters}
We report results based on the simulations listed in \autoref{tab:example_table}, performed using the \Abacus $N$-body code \citep{Garrison2021b}. \Abacus offers high performance and accuracy, based on a high-order multiple method to solve far-field forces and an accelerated GPU calculation of near-field forces by pairwise evaluation. While the $N=1024^3$ simulations were run using local facilities at the Harvard-Smithsonian Center for Astrophysics (CfA), the larger $N=4096^3$ simulations are part of the \textsc{AbacusSummit} project \citep{Maksimova2021}, which used the Summit supercomputer of the Oak Ridge Leadership Computing Facility.

The simulation data we exploit in this article are also summarized in \autoref{tab:example_table}. As in \citet{Maleubre2022}, we have simulated three different exponents ($n=-1.5$, $n=-2.0$ and $n=-2.25$), chosen to probe the range relevant to standard (i.e. LCDM-like) models. 
For the first two exponents, we have two simulations with different $N$ but otherwise identical parameters, allowing us to study the impact coming from finite box size effects. 
For the larger ($N=4096^3$) simulations, the statistics have been 
calculated on (random) sub-samples of different sizes (25\%, 3\%) to facilitate the assessment of finite sampling effects. 
For the other two spectral indices, $n=-2.0$ and $n=-2.25$, we have four 
$N=1024^3$ simulations, each with identical $N$-body parameters but
different realizations of the IC. These will be analysed below, both 
individually and as an average, to complement the finite sampling vs. finite box size effects analysis. 

Thorough this whole study, we work in units of the mean inter-particle (i.e. initial grid) spacing, $\Lambda=L/N^{1/3}$. The essential time-stepping parameter in
\Abacus has been chosen as $\eta=0.15$ for all simulations, and 
the additional numerical parameters have been set as detailed in \citet{Maleubre2022} and summarized below. These choices are based on the extensive convergence tests of these parameters reported in our previous studies \citep[see][]{Joyce2021,Garrison2021}.

The start of the simulation ($a=a_i$) is chosen so that top-hat density fluctuations at the particle spacing are given by
\begin{equation}
    \sigma_i(\Lambda,a_i)=0.03
\end{equation}
while the first output epoch ($a=a_0$)\footnote{We emphasize that for this paper $a_0$ corresponds to the first output of the simulation, and not the scale factor today, as it's usually the case.} corresponds approximately to the formation of the first non-linear structures, fixed at the time at which fluctuations of peak-height $\nu \approx 3$ are expected to virialize in the spherical collapse model ($\sigma \sim \delta_{c}/\nu$, with $\delta_{c}= 1.68$):
\begin{equation} \label{eq:def_a0}
\sigma_{\rm lin}(\Lambda,a_0)=0.56
\end{equation}

Subsequent output values are spaced by a factor $\sqrt{2}$ in the non-linear mass scale. Given that $M_{\rm NL}\propto R_{\rm NL}^3$ and substituting in \autoref{eq:RNL-scaling}, we get:
\begin{equation}\label{eq:diff_a}
    \Delta\log_2 a = \frac{3+n}{6}\Delta\log_2 M_{\rm NL}=\frac{3+n}{12} 
\end{equation}

We use $\log_2(a/a_0)$ as the time variable of our analysis, which indicates how many epochs have passed since the first output. It is also convenient to define the variable
\begin{equation}\label{eq:Sdef}
    S = \frac{12}{3+n}\log_2\left(\frac{a_S}{a_0}\right)
\end{equation}
with $S=0,1,2,...$ corresponding to the different outputs of the simulation. 

With respect to the force softening, as previously described in \citet{Garrison2016}, \Abacus uses a spline softening derived using a Taylor expansion to second order in $r$ of a Plummer softening expression, taking the form
\begin{equation}
        \mathbf{F}(\mathbf{r})=
    \begin{cases}
        \left[10-15\left(r/\epsilon_s\right)+16\left(r/\epsilon_s\right)^2\right]\mathbf{r}/\epsilon_s^3, & r<\epsilon_s \\
        \mathbf{r}/r^3, & r\geq \epsilon_s
    \end{cases}
\end{equation}
which imposes a smooth transition at the softening scale $\epsilon_s$. All softening lengths in this study have been fixed in proper coordinates, decreasing as $\epsilon(a)\propto 1/a$ in comoving coordinates, those used by the simulation. To avoid a too large softening at earlier times, we first fixed it in comoving coordinates down to $a_0$, the first output of our simulation, and change it to proper from then on. For all the simulations studied here, we use $\epsilon(a_0)/\Lambda=0.3$. This value has been chosen following the results in \citet{Garrison2021} and \citet{Maleubre2022}, being both accurate and efficient for the spectral indices analysed.

Initial conditions have been set up using a modification to the standard Zel'dovich approximation (ZA), detailed in \citet{Garrison2016}. This includes a second order Lagrangian perturbation theory (2LPT) correction as well as particle linear theory (PLT) corrections as described in \citet{Joyce2007} and \citet{Garrison2016}. The latter corrects the initial conditions for discreteness effects at early times, so that the result of fluid evolution is reproduced at a target time 
$a=a_{\rm PLT}$. For all our simulations here we have $a_{\rm PLT}=a_0$, with $a_0$ defined by \autoref{eq:def_a0}.

\begin{table*}
	\centering
	\caption{Summary of the $N$-body simulation data used for the analysis of this paper. The first column shows the spectral index of the initial PS, $N$ is the number of particles of each simulation, and the third column shows the number of simulations with identical parameters but different realizations of the IC. The fourth column shows the available statistic and sampling of the matter field.}
	\label{tab:example_table}
    \begin{tabular}{cccccc}
    \hline
    $n$         & $N$         & num. sims. & DM Statistic ($\%$)    \\ \hline \hline
    $n=-1.5$  & $4096^3$  & 1          & $v_r$ and $\xi$ (25\%)     \\
    $n=-1.5$  & $1024^3$  & 1          & $v_r$ and $\xi$ (100\%)    \\ \hline
    $n=-2.0$  & $4096^3$  & 1          & $\xi$ (3\%)                \\
    $n=-2.0$  & $1024^3$  & 4          & $v_r$ and $\xi$ (100\%)    \\ \hline
    $n=-2.25$ & $1024^3$  & 4          & $v_r$ and $\xi$ (100\%)    \\ \hline
    \end{tabular}
\end{table*}

\subsection{Estimation of converged values}\label{sec:converge_method}

As in our previous papers, we will assess the convergence to the physical limit of a particular statistic by studying its temporal evolution, which becomes time-independent in rescaled variables in the case of self-similarity. To make this study quantitative --- i.e. to identify estimated converged values, and converged regions at some precision --- we need to adopt appropriate criteria.
While the conclusions drawn should not of course depend significantly on the chosen criteria, these criteria are intrinsically somewhat arbitrary in detail. In practice, their choice is made based on visual examination of data. We follow here the simple procedure described in \citet{Maleubre2022}. It allows us to estimate a converged value and converged region at a chosen precision, per rescaled bin for each of the statistics analysed in this paper. The method is equivalent for all our matter-field dimensionless statistics($\xi$, $\bar{\xi}$, $v_r/Hr$), and we denote our chosen statistic by $X$ in the following explanation.

Studying temporal self-similar evolution at determined rescaled bins consists in analysing the behaviour of the statistic at vertical slices in the right panels of \autoref{fig:SS_SF}.

We first calculate an estimated converged value (denoted as $X_{\text{est}}$) in each rescaled bin as the average of the statistic in a specific temporal window. The width of this window 
is conveniently specified by a number of snapshots $w$, corresponding to an increase in the 
non-linearity scale by a factor\footnote{Remember that for each subsequent snapshot the non-linearity scale increases by $2^{1/6}$, so the total increase over a time window $w$ will be $2^{w/6}$.} of $2^{w/6}$ (below we use $w=5$). To identify which time-window in the span of the whole simulation is the best converged, we ``slide'' our window of width $w$ across the data to find the one which minimizes
\begin{equation}
    \Delta=\frac{\left|X_{\text{max}}-X_{\text{min}}\right|}{2\mu_{X}}
\end{equation}
where $X_{\text{max}}$, $X_{\text{min}}$, and $\mu_{X}$ are respectively the maximum, minimum, and average values in the window. 
As a result, $X_{\text{est}}$ is a first attempt to calculate the most self-similar value of the requested statistic in a particular rescaled bin. But in reality, we're only interested in convergence above a particular precision. Specifying now a parameter $p$ characterizing the precision of convergence, any bin is considered to be converged only if the minimal value of $\Delta$ is less than $p$. 

As a result, we now know which rescaled bins are converged at a precision $p$ or better, for our studied statistic. Ultimately, we're interested in identifying the minimum physical scales at which we have access to at any given time. For this, we now need to identify the maximum temporal window behaving self-similarly for each of the converged rescaled bins.

To identify the temporal region of convergence with respect to $X_{\text{est}}$ (still at precision $p$), for each rescaled bin we find the largest (containing at least three consecutive snapshots, though again this number is not essential) connected temporal window verifying
\begin{equation}\label{eq:X_conv}
    \frac{|X-X_{\text{est}}|}{X_{\text{est}}}<p\,.
\end{equation}
We denote $X_{\text{conv}}$ the average calculated over this new largest temporal 
window, and take this as the final estimated converged value of the statistic 
for the given rescaled bin. The edges of the window are thus the earliest and latest converged snapshots.
We note that, in the following \citep[as in][]{Maleubre2022}, when we say 
that we have precision at $x\%$ we mean that $p=x/100$~\footnote{What we denote $p$ here corresponds to $\alpha/2$ in \citet{Maleubre2022}.}. 

In the results presented below, all two-point quantities have been calculated over the same $r/R_{\rm NL}$ grid. We use bins of constant logarithmic spacing $1+(\Delta r/r)\approx2^{1/12}$ \citep[][following]{Maleubre2022}, ensuring that bins of different snapshots match when rescaled by $R_{\rm NL}$ to facilitate comparison between them. In order to reduce statistical noise sufficiently, we have rebinned by grouping four such bins, corresponding to $\Delta r/r\approx0.26$. In our presentation below we label our bins, for simplicity, just by the value of the rescaled variable at the geometrical centre of the bin.

%% file: Sections/4_Results.tex

As discussed above, in a scale-free cosmology, self-similarity implies an independence of the results of an $N$-body simulation of their discretization parameters. By carefully examining the departures from self-similarity that are actually measured, we can infer how the resolved scales depend on the
unphysical scales in the $N$-body simulation. We report in this section this analysis for the mean pairwise velocity in the matter field.

\subsection{Direct estimation}

As discussed in \autoref{sec:PW_PC}, $v_{r}/Hr$ can be estimated directly from the measured particle velocities, or indirectly from measurements of the 2PCF. We consider first the former estimate. \autoref{fig:SS_SF} shows the estimated $v_{r}/Hr$ as a function of time (parameterized by the variable $\log_2(a/a_0)$) at different distances, for spectral indices $n=-1.5$, $n=-2.0$ and $-2.25$. Each plot correspond to the simulations with the highest number of particles ($N=4096^3$ for $n=-1.5$,  and the average of the four $N=1024^3$ simulations for the other). The left panel gives $v_r/Hr$ as a function of $r/\Lambda$ (with $\Lambda$ the grid spacing), while the right panel gives it as a function of the rescaled variable $r/R_{\rm NL}$. Self-similarity corresponds to the superposition of the data at different times in the latter plot.

These plots show qualitatively the general behaviour of the statistic, which is similar to that seen for the 2PCF \citep{Joyce2021} and the PS \citep{Maleubre2022}.
Self-similarity can be seen to propagate from larger comoving scales, significantly above $\Lambda$ at early times, to smaller scales as time evolves. 
In particular, the scales around the ``turnaround'' point --- corresponding to the maximal radial infall velocity --- are only resolved at later times. 
As for the 2PCF and PS in our previous studies, the redder the index, the more reduced is the range of approximate self-similarity. This is a reflection primarily of the smaller range of scale-factor which is accessible in simulations of a fixed size as $n$ decreases, and also, as we will see further below, of larger finite box size effects.
Finally, we note that all three models appear to show the same behaviour at 
asymptotically small scales, tending to a value close to $-1$, the  value predicted by the stable clustering hypothesis. We will assess these behaviours 
quantitatively  below in \autoref{sec:ConvSS}.

\subsection{Estimation using pair conservation}

\begin{figure*}
  \begin{subfigure}{\linewidth}
    \includegraphics[width=\linewidth]{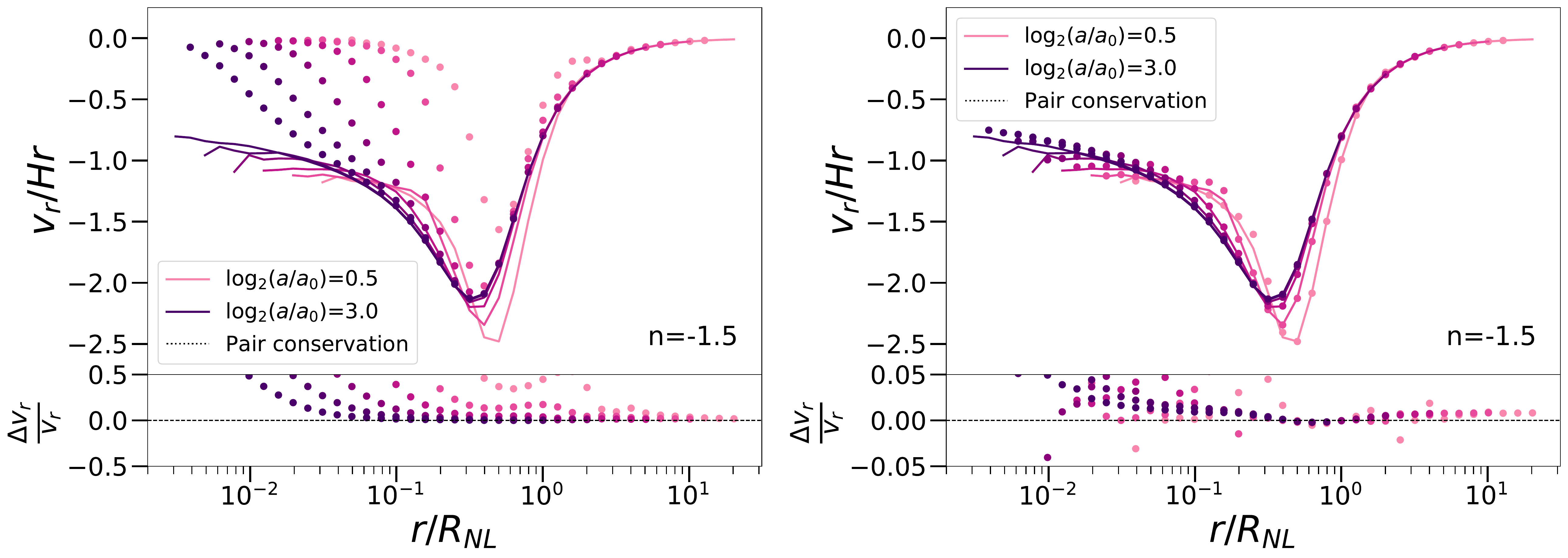}
  \end{subfigure}
  \begin{subfigure}{\linewidth}
    \includegraphics[width=\linewidth]{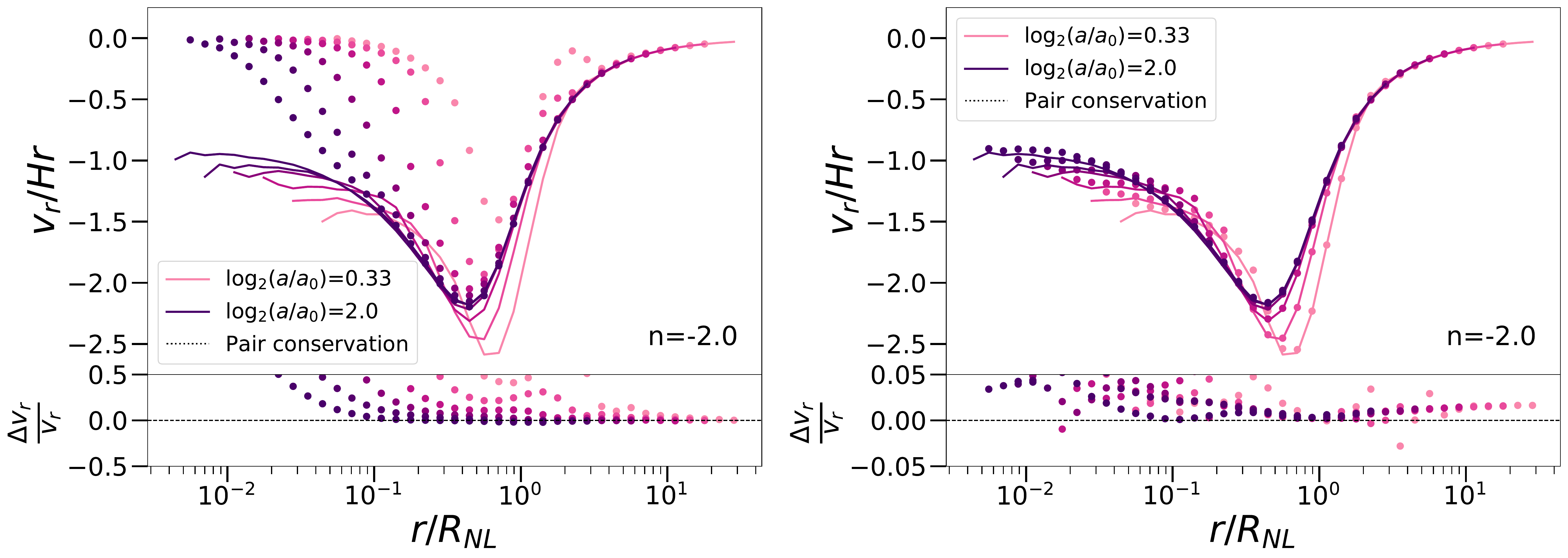}
  \end{subfigure}
  \begin{subfigure}{\linewidth}
    \includegraphics[width=\linewidth]{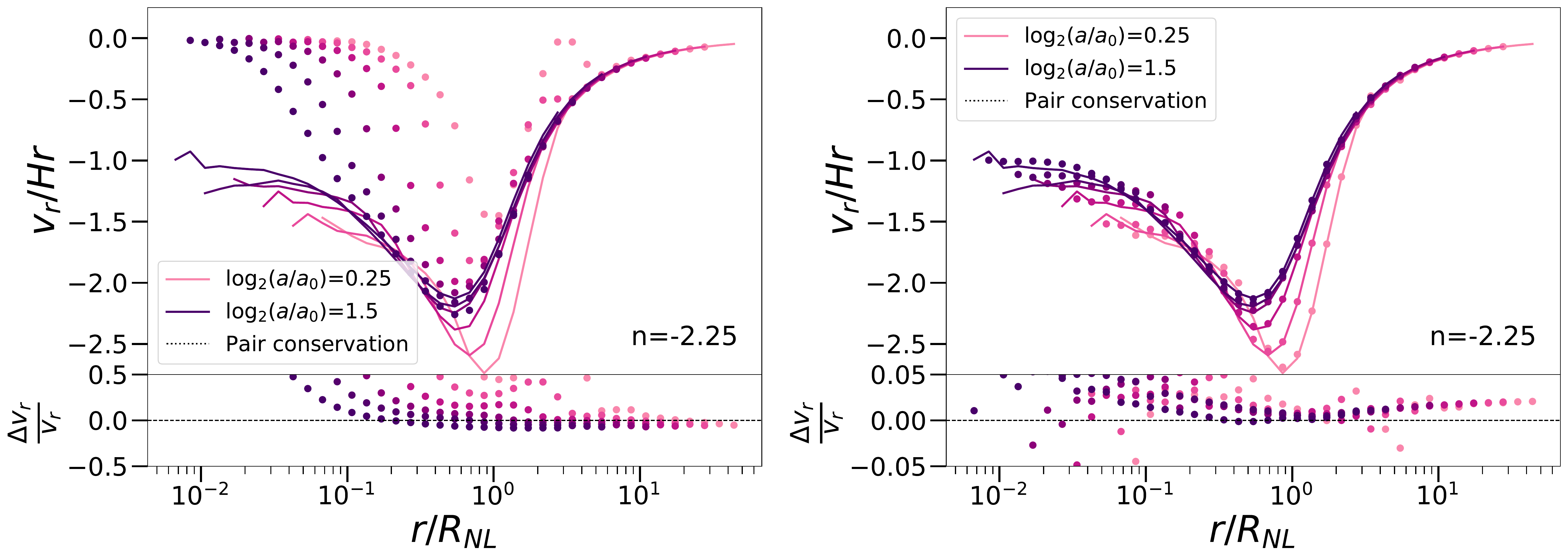}
  \end{subfigure}
  \caption{Comparison of different estimators of $v_r/Hr$ as a function of the rescaled length $r/R_{\rm NL}$, for the same simulations as in \autoref{fig:SS_SF}.
  The  solid lines in each pair of panels (left and right) are identical and correspond to the results obtained by direct estimation using the velocities (as in \autoref{fig:SS_SF}). The dots correspond, in the left panels, to estimations using pair-counting and the assumption of self-similarity of the 2-pt statistics
  i.e. using  \autoref{eq:pw_pc} with the last term set to zero. In the right panels, this last term is also included in the estimator. The bottom section of each panel shows the difference between pair conservation estimation and direct estimation, with respect to the latter. Note that the y-axis scale has been reduced by $x10$ from the left to the right panels.}
  \label{fig:PW_PC}
\end{figure*}

We next consider the estimation of $v_r$ from the 2PCF, using the exact relation \autoref{eq:pw_pc} for $v_r$ in terms of $\xi$, $\bar{\xi}$ and $\dot{\bar{\xi}}$. As noted, we can also test the validity of the relation when the term in $\dot{\bar{\xi}}$ vanishes, which corresponds to self-similarity of $\bar{\xi}$. 
\autoref{fig:PW_PC} shows the normalized pairwise velocity at each rescaled coordinate for a set of selected redshifts, in the same way as in the right panel of \autoref{fig:SS_SF}. In addition, we have added a dotted line which gives the new estimation obtained using pair conservation. The left panel excludes the 
non-self-similar term,  while the right panel corresponds to the full (exact) expression \autoref{eq:pw_pc}. Finally, each of the small sub-panels shows the scatter between results from direct estimation and pair conservation.
To estimate the time derivative, we have simply used a finite difference estimate on the closest two ``neighbouring'' snapshots. 

In the right panels we see that, as required by pair conservation, we recover $v_r$ to a very good approximation from the alternative estimator. The very small differences can be attributed to finite particle number noise and possible systematic offsets due to the discrete estimation of the time derivative. Given the close spacing (\autoref{eq:diff_a}) of our snapshots, it is unsurprising that any such effect appears to be small. At small scales, on the other hand, close examination shows that the pair conservation estimator is slightly less noisy than the direct one. This is as might be anticipated: because of the intrinsic dispersion in the pairwise velocities, we can expect its average to have a greater variance than the direct pair count \citep[as noted previously by][]{Jain1997}. Thus, in assessing what is required to obtain an accurate estimation of
the pairwise velocity, one needs to consider between the need to have closely spaced outputs to accurately estimate the time derivative if pair counting is used, or a larger volume for accurate direct estimation.

\begin{figure}
  \begin{subfigure}{\linewidth}
    \includegraphics[width=\linewidth]{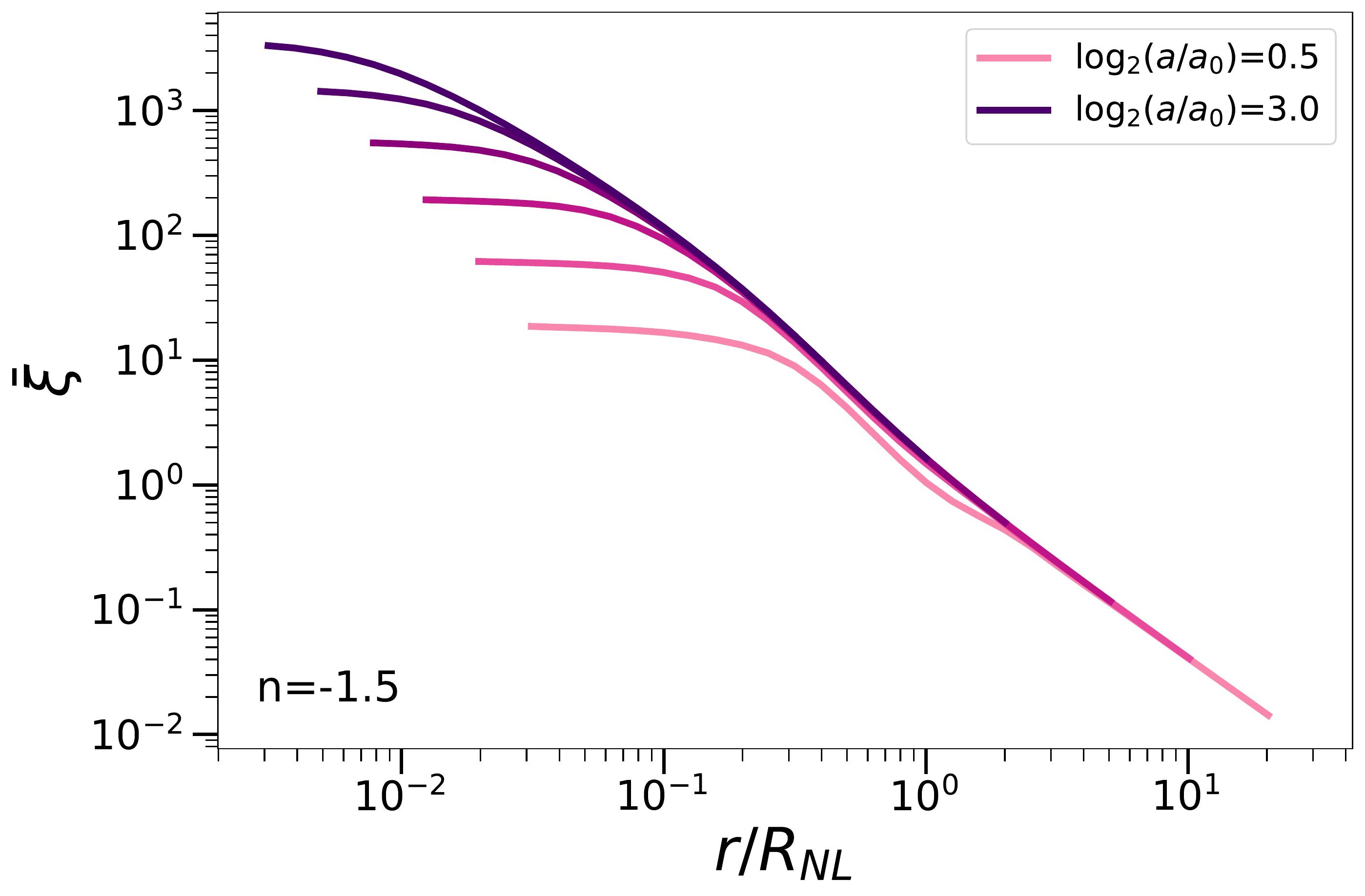}
  \end{subfigure}
  \begin{subfigure}{\linewidth}
    \includegraphics[width=\linewidth]{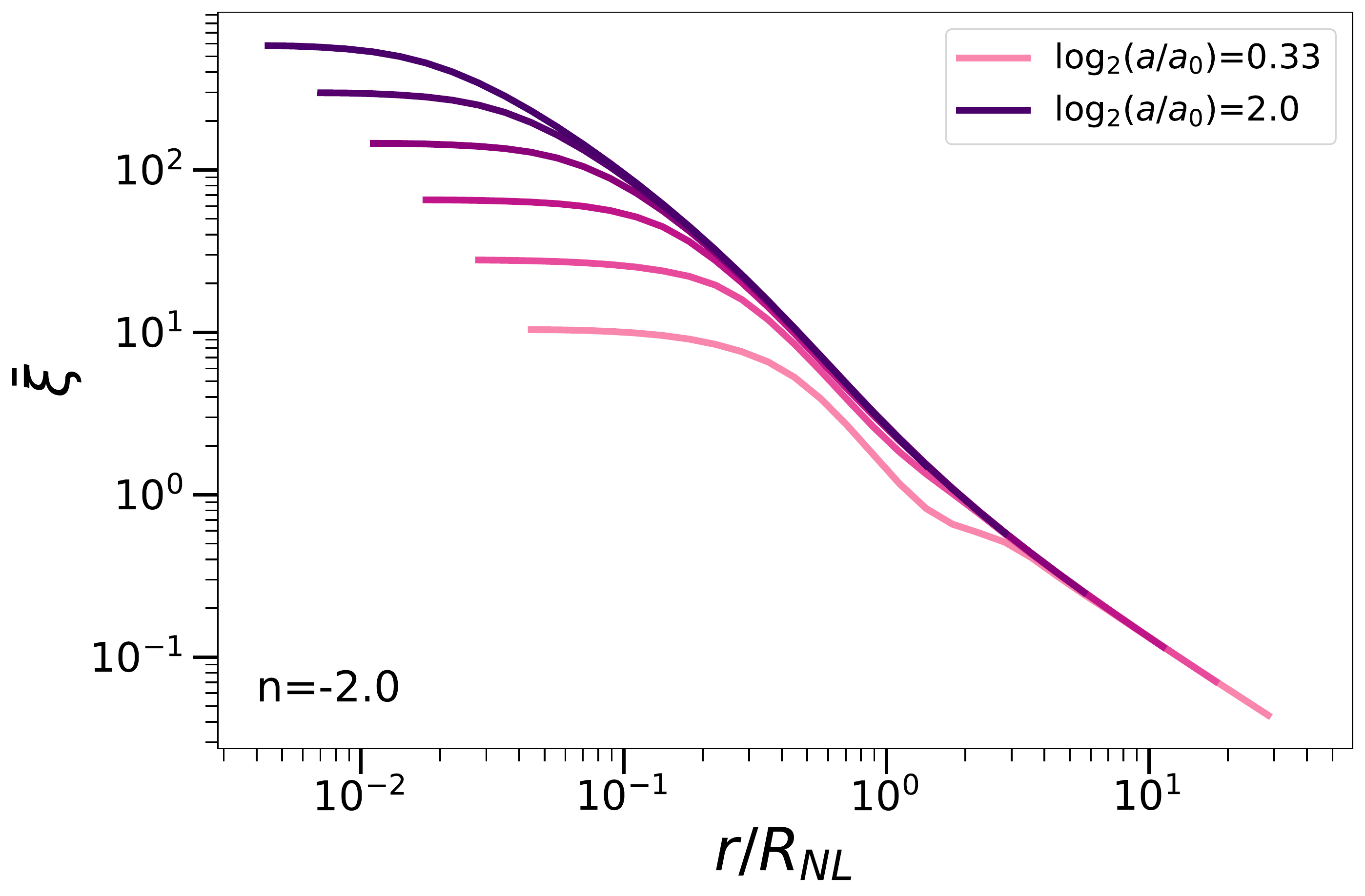}
  \end{subfigure}
  \begin{subfigure}{\linewidth}
    \includegraphics[width=\linewidth]{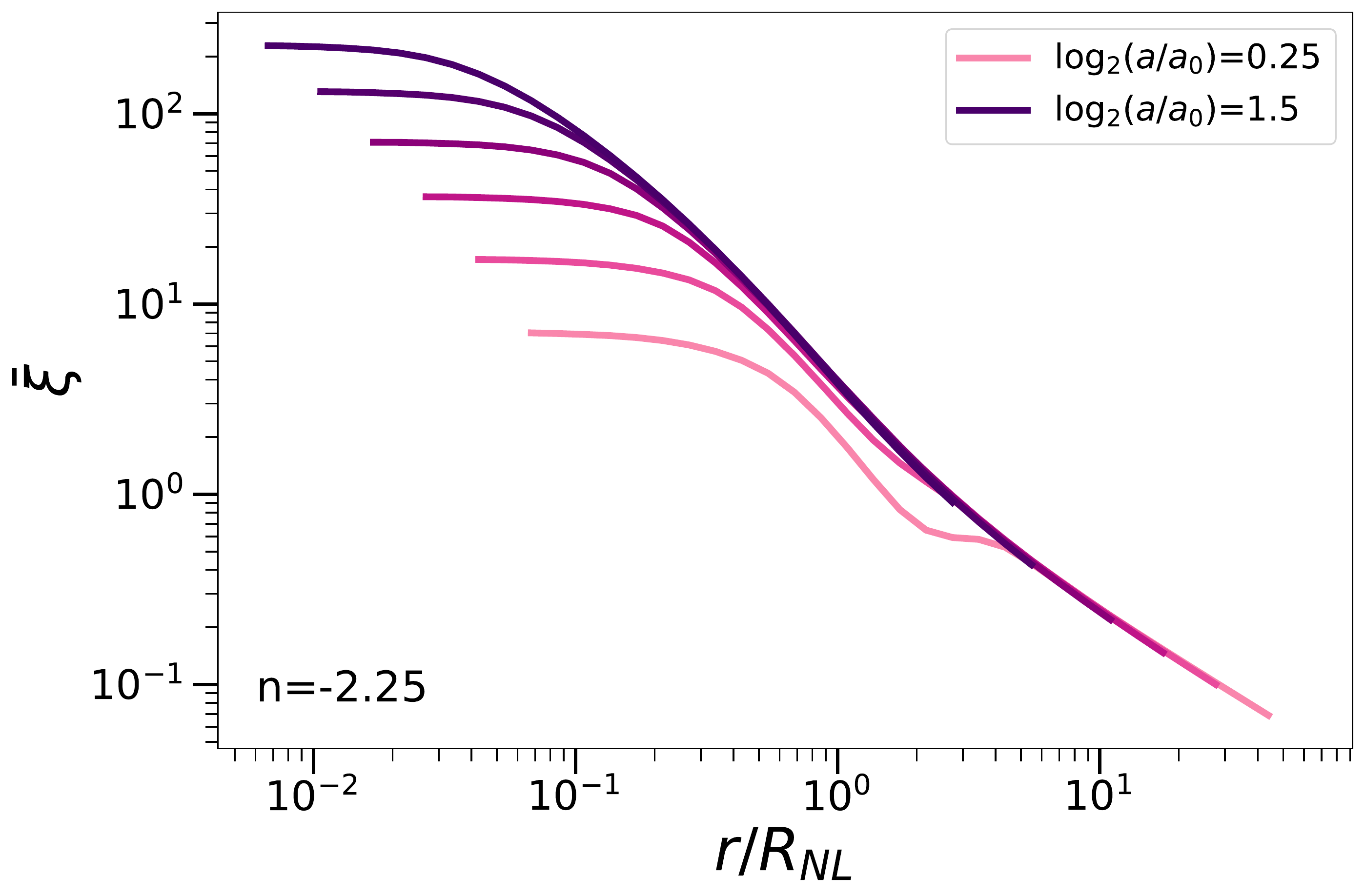}
  \end{subfigure}
  \caption{\emph{Cumulative} 2PCF as a function of rescaled length $r/R_{\rm NL}$ at same times as in \autoref{fig:PW_PC}. Simulations correspond to $n=-1.5$, $n=-2.0$ and $n=-2.25$, with $N=4096^3$ in the former exponent and the average over four $N=1024^3$ for the other two.}
  \label{fig:c2PCF}
\end{figure}

The left panels, on the other hand, show very large discrepancies between the two estimators, which we can infer as being due to a significant deviation from self-similarity in the corresponding range of the (integrated) 2PCF. Indeed, we can see that this is the case from the corresponding direct analysis of $\bar{\xi}$ displayed in \autoref{fig:c2PCF}: the scales at which the agreement of the estimators break corresponds to the break from self-similarity of 
$\bar{\xi}$. We note that, at late times, the associated break appears to occur at a scale where $v_r/Hr$ approaches $-1$, the value corresponding to stable clustering.
Thus, there is indeed a range where approximate self-similarity in $v_r$ appears to persist 
despite the fact that the 2PCF differ much more from their physical values, and this range appears to correspond, at later times, to that where stable clustering is well approximated.

\subsection{Quantitative determination of resolved scales}\label{sec:ConvSS}

To better understand, and then also quantify, the limitations on the range of self-similarity arising from the different unphysical simulation parameters (specifically $\Lambda$, $\epsilon$ and $N$) we now study more closely the evolution as a function of time of $v_r/Hr$ (estimated directly and indirectly via pair conservation), and of $\xi$ and $\bar{\xi}$, for fixed values of $r/R_{\rm NL}$. This corresponds to taking the values on vertical lines in the right panels of \autoref{fig:SS_SF} (and the equivalent plots for $\xi$ and $\bar{\xi}$). As discussed, self-similarity of the statistic then corresponds to time independence, i.e. to convergence (in some range) of the time series to a fixed value. 

\begin{figure*}
  \begin{subfigure}{\linewidth}
    \includegraphics[width=\textwidth]{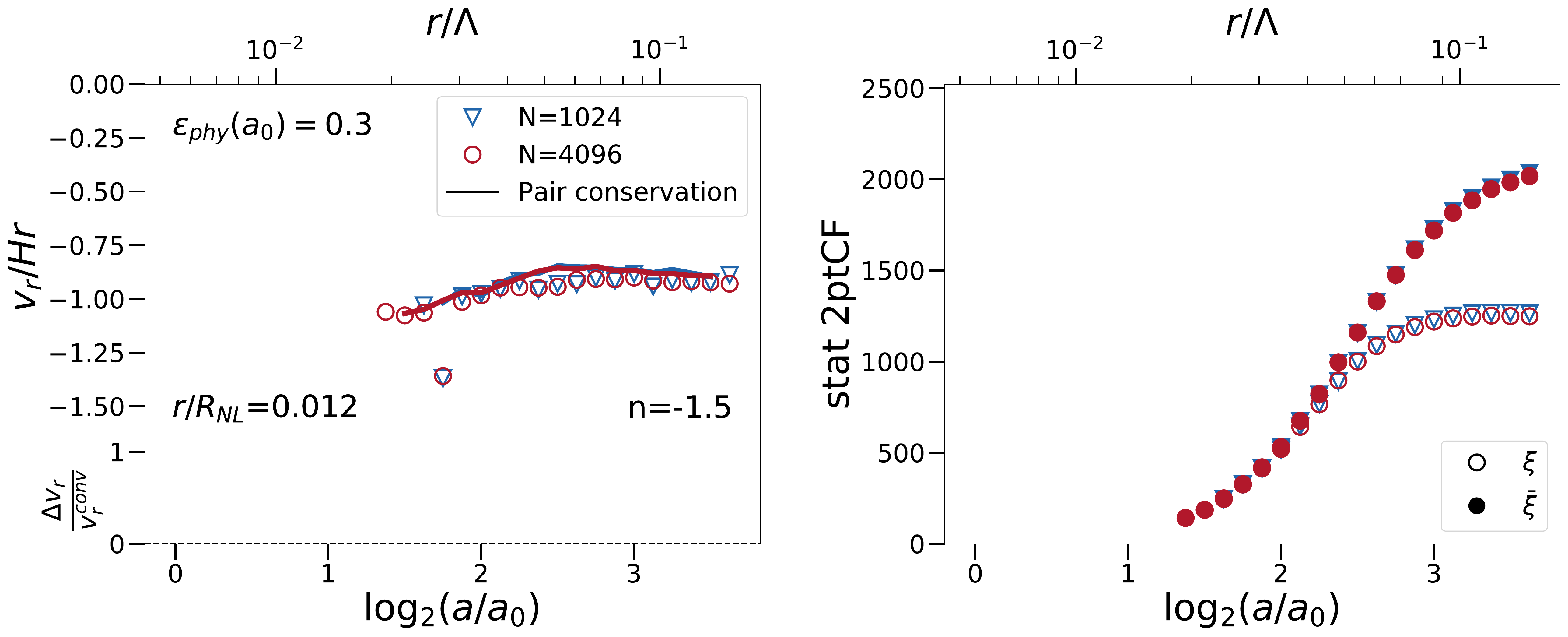}
  \end{subfigure}
  \begin{subfigure}{\linewidth}
    \includegraphics[width=\textwidth]{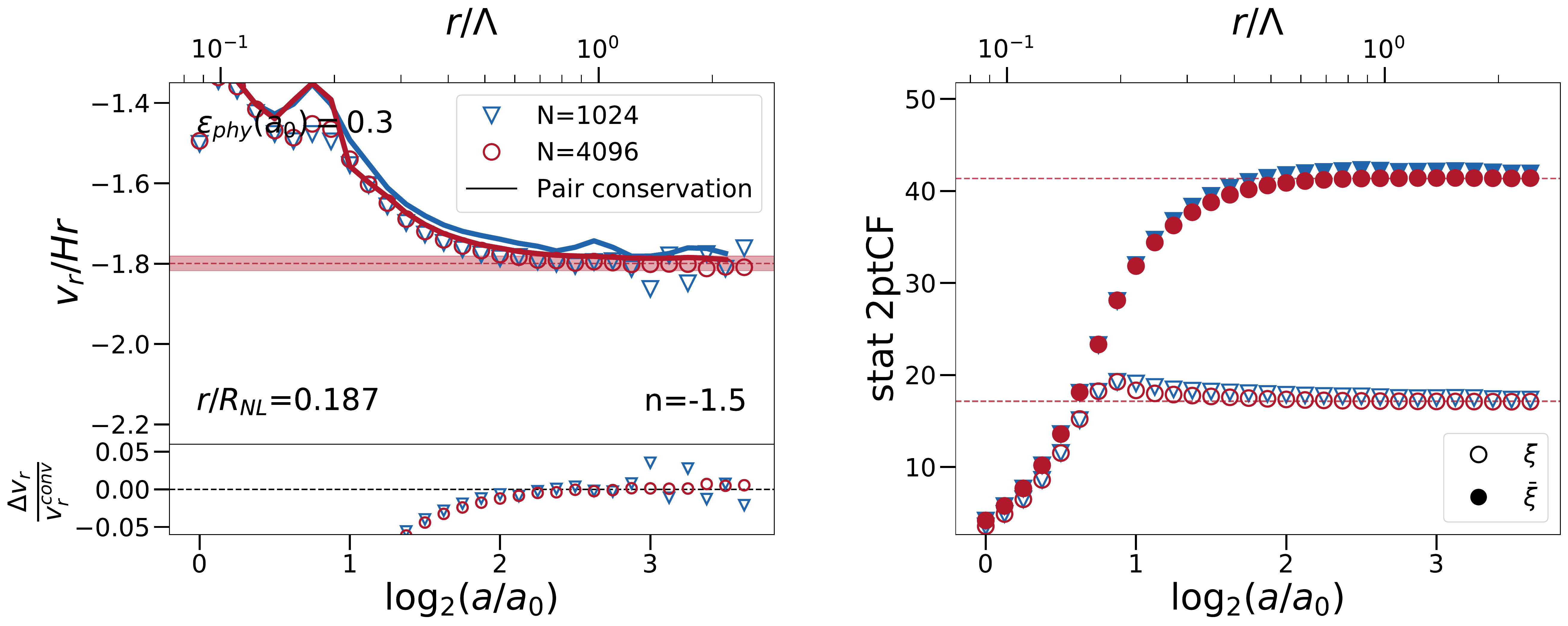}
  \end{subfigure}
  \begin{subfigure}{\linewidth}
    \includegraphics[width=\textwidth]{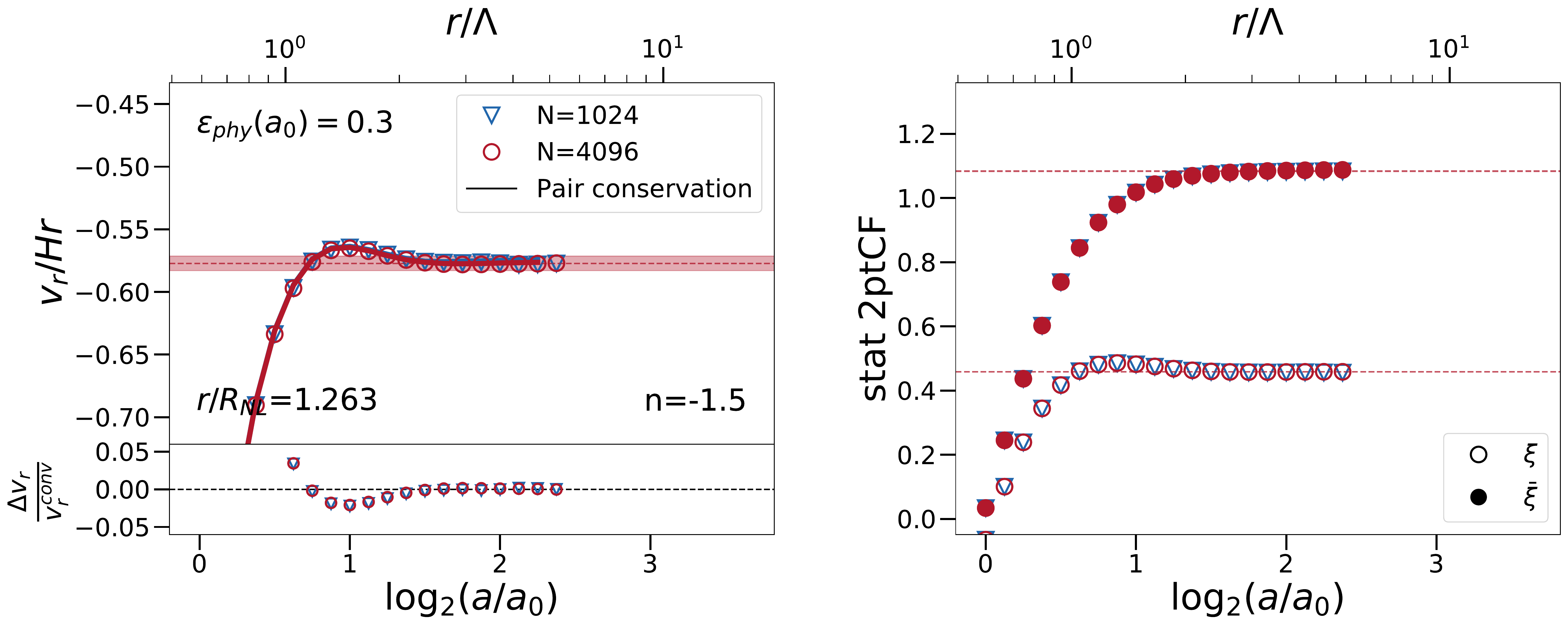}
  \end{subfigure}
  \caption{
  Evolution, for $n=-1.5$ simulations, of $v_r/Hr$ (left panels), and of the 2PCF and \emph{cumulative} 2PCF (right panels), as a function of logarithmic scale factor $\log_2(a/a_0)$, \emph{lower x-axis}, and as a function of $r/\Lambda$, \emph{upper x-axis}. 
  Each row correspond to a different bin of rescaled separation $r/R_{\rm NL}$ as labelled. 
  The blue triangular symbol represents the smaller $N=1024^3$ simulation, while the red circles represent the $N=4096^3$ simulation. Results obtained using the pair counting estimator are drawn as a continuous line in the appropriate colours.  Horizontal red dashed lines indicate the converged value of each of the three statistics, calculated from the largest simulation as described in the text, and the  red shaded region indicates that within $\pm1\%$ of this value. The sub-panels in the plots of $v_r/Hr$ give the dispersion of the results obtained using the direct estimation with respect to the converged value.}
  \label{fig:SS_n15}
\end{figure*}

\begin{figure*}
  \begin{subfigure}{\linewidth}
    \includegraphics[width=\linewidth]{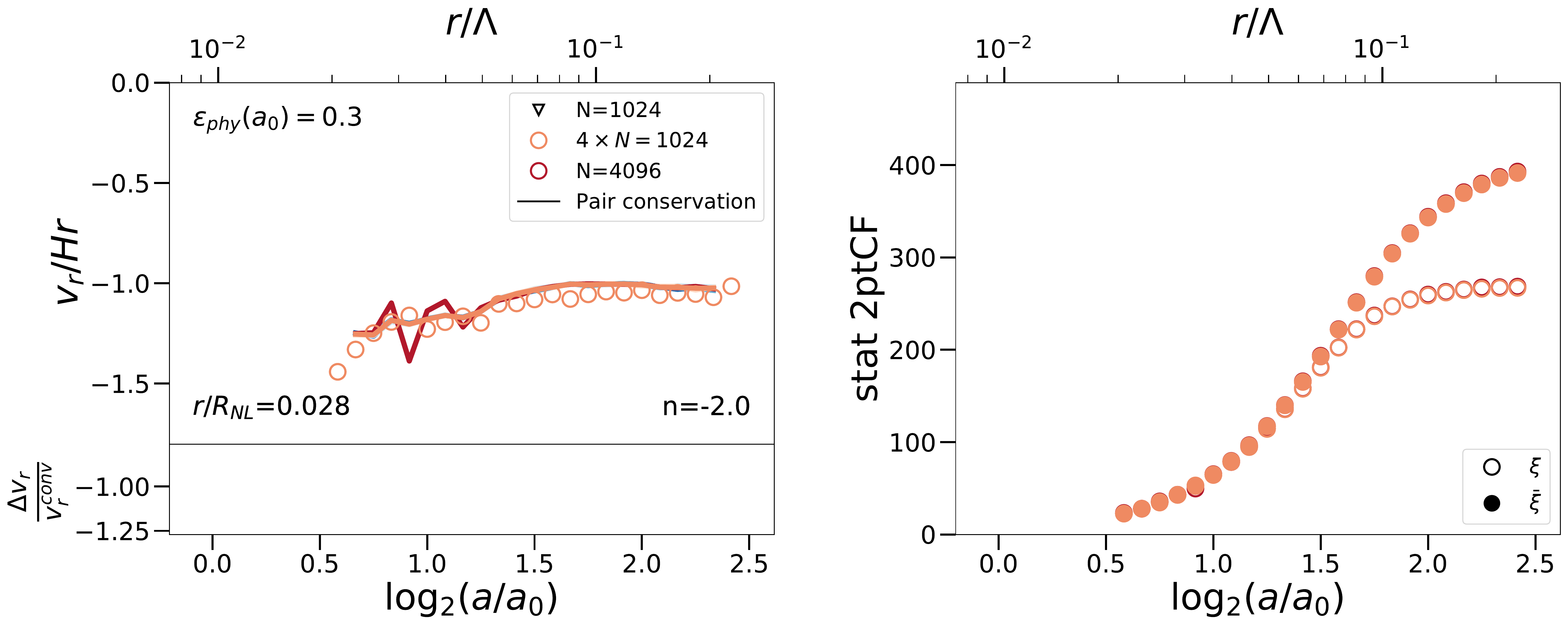}
  \end{subfigure}
  \begin{subfigure}{\linewidth}
    \includegraphics[width=\linewidth]{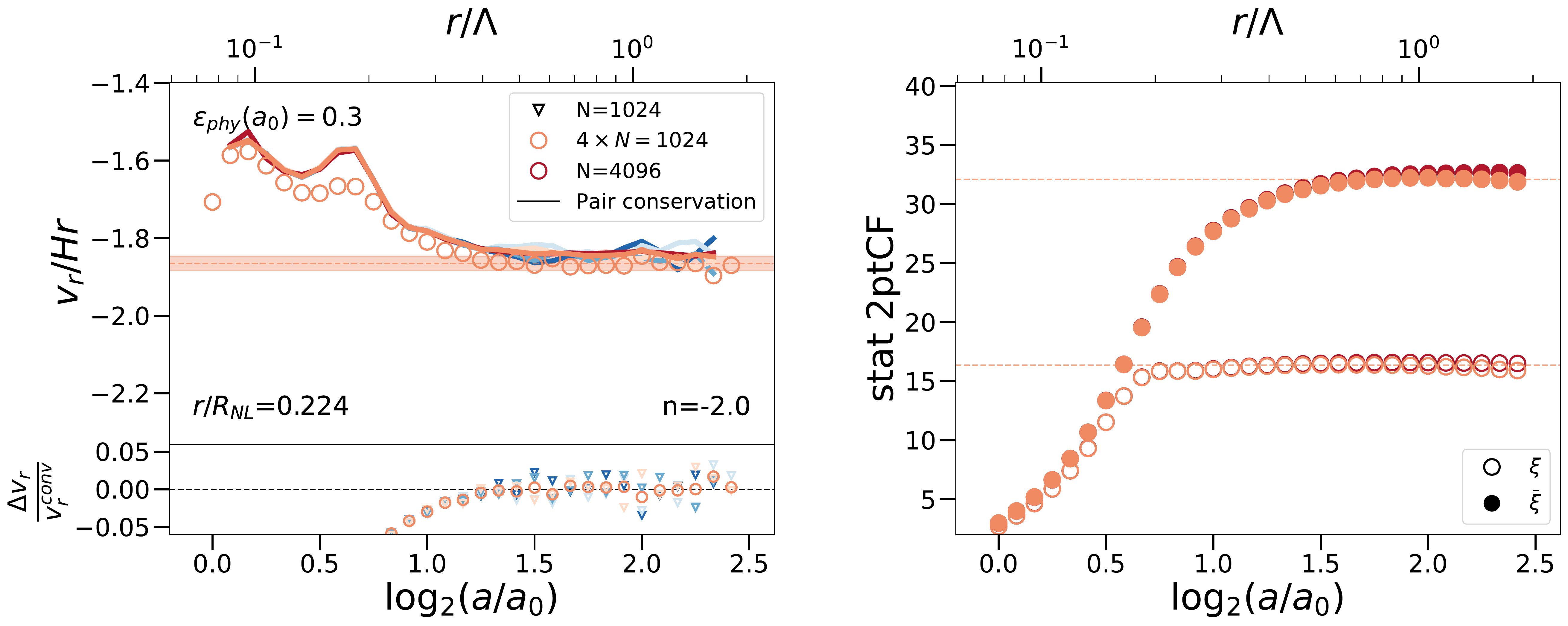}
  \end{subfigure}
  \begin{subfigure}{\linewidth}
    \includegraphics[width=\linewidth]{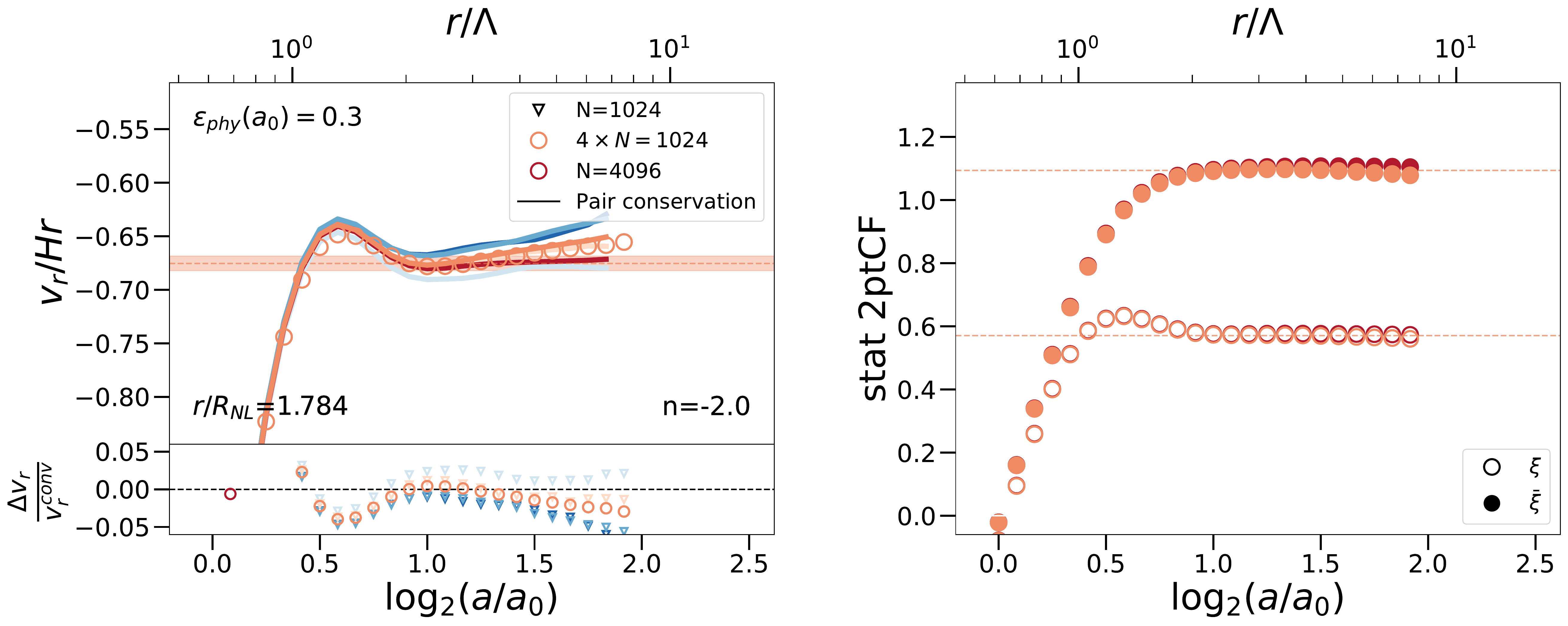}
  \end{subfigure}
  \caption{Same as \autoref{fig:SS_n15}, but for $n=-2.0$ simulations. There are now four sets of triangular symbols representing the different $N=1024^3$ simulations (in the sub-panels only), while the orange circles represent the average of the four. Note further that the results for $v_r/Hr$ from the $N=4096^3$ simulation are all obtained by pair counting only, so that there are no red circles in the left panels. 
}
  \label{fig:SS_n20}
\end{figure*}

\autoref{fig:SS_n15} and \autoref{fig:SS_n20},  
for spectral indices $n=-1.5$ and $n=-2.0$ 
respectively, shows such plots for three chosen values of $r/R_{\rm NL}$. (We exclude n=-2.25 for economy, but will discuss it further below). To help understand the scales involved in each plot, we also display the values of $x/\Lambda$ on the upper $x$-axis. As $R_{\rm NL}$ is a monotonically growing function of time, $x/\Lambda$ increases from left to right, translating the fact that the spatial resolution relative to the grid increases with time in these plots. We note that in almost all the plots we can identify easily by eye what appears to be a converged value in a finite range of scales (the only exceptions are those of
$\bar{\xi}$ in the first panels). In all these cases, a lower cut-off to this converged 
range is clearly identifiable. 
As we discussed in the analysis of similar plots in our previous analyses \citep{Joyce2021,Maleubre2022}, and will see again in detail now, this lower cut-off clearly corresponds to the resolution limit fixed by the ultraviolet cut-offs ($\Lambda$ and $\epsilon$).

The different estimators of the statistics shown are indicated in the legend and described in the figure caption. Recall that, as detailed in \autoref{tab:example_table}, the properties of the simulations analysed differ for the two different exponents. While data for $n=-1.5$ correspond to a single realization of each box size, $n=-2.0$ presents data from four different realizations of $N=1024^3$ boxes and their statistical average.

In the cases in which the rescaled bin is converged following the criterion specified above in \autoref{sec:converge_method}, at a precision of $1\%$ (i.e. $p=0.01$), the estimated converged value is indicated as a dashed line and the red shaded region indicates that within $1\%$ of this value. We only plot the converged value of the biggest simulation (for n=-1.5) and the one coming from the averaged statistic (for n=-2.0), but visual inspection can help assess the convergence of the other boxes, whose converged value needs to coincide.
In addition, and to help the reader evaluate the degree of convergence of the different boxes, we add a sub-plot with the dispersion between this converged value and the individual data points from direct estimation coming from \emph{all} our simulations (including the individual $N=1024^3$ boxes with $n=-2.0$). This value of $1\%$ is chosen because it is approximately the smallest value of $p$ for which we obtain a significant range of contiguous bins satisfying our convergence criteria. It corresponds to the highest precision (i.e. smallest $p$) at which we can in practice establish convergence using our data.

The first panel of each figure corresponds to a highly non-linear (small) scale.  Although $v_r/Hr$ is not converged at the $1\%$ precision level, the different estimators nevertheless give highly consistent values and appear to show robust convergence albeit at lower precision (of order a few percent), starting from a scale well below $\Lambda$. As anticipated in the previous section, the converged value is close to $-1$. Further, we see more clearly that this convergence is indeed not associated with that of $\bar{\xi}$, i.e. at this scale the measured \emph{cumulative} 2PCF $\bar{\xi}$ approximates very poorly its physical value.

The next (second) set of panels (of both \autoref{fig:SS_n15} and \autoref{fig:SS_n20}) corresponds to the bin around the smallest rescaled separation for which $v_r/Hr$ (in the statistically largest available simulation, using direct estimation) converges (according to our convergence criterion, at the chosen $1\%$ precision level). The lower cut-off to the convergence of $v_r/Hr$ is just slightly below the grid spacing (at about $\Lambda/2$). We see also that $\bar{\xi}$ shows convergence starting from the same scale, so the range of convergence for the pair counting estimator using $\dot{\bar{\xi}}=0$, i.e. assuming self-similarity of $\bar{\xi}$, will be accurate in a similar range.
Looking at the lower sub-panels in the plots of $v_r/Hr$, we see that the convergence of the direct estimators in the individual $N=1024^3$ simulations is 
degraded just above $\Lambda$ for $n=-1.5$ and slightly below for $n=-2.0$. This is simply finite $N$ noise in the estimators, as
the associated fluctuations disappear in the larger ($N=4096^3$) simulation for $n=-1.5$ but also when the four $N=1024^3$ simulations are combined for $n=-2$ (thus ruling out finite boxsize as the origin of these differences).

The third row of both figures shows a considerably larger scale, in the weakly non-linear regime, which have a lower cut-off to convergence (again, at the $1\%$ level) a few times larger than the grid spacing. In this case, for $n=-1.5$, there is no visible evidence for the finite $N$ effects seen in the previous bin. On the contrary, for $n=-2.0$, we observe much poorer convergence of $v_r/Hr$, both in the direct estimations (symbols) and in the pair counting estimator (solid lines in main panel), and it is visible whether we look at the individual simulations or the average statistic. Further, we see now an offset from the estimated converged value that is a systematic shift rather than a random noise, and even in the average over the four simulations. Thus, we can detect a break from convergence within the range of scales probed. The cancellation (or at least partial cancellation) of these systematic offsets when the realizations are averaged indicates that, at intermediate scales, this is due to significant differences in the initial power at larger scales due to the finite sampling of modes. On the other hand, the observed break from convergence at larger scales (in the average) can be attributed to finite box size effects arising from the missing power in modes below the fundamental of the simulation box, finite $L$, and no longer due to a finite $N$ as before. These same tendencies are present, but even much more pronounced for $n=-2.25$ (data not shown). Indeed, in this case, the lower and upper cut-offs to convergence below the few percent level are no longer clearly separable from one another in almost all bins. For this reason, we do not use the $n=-2.25$ below in our quantitative assessment of resolution limits.

Summarizing, we can state that:
\begin{itemize}
    \item At any given converged scale, direct estimation and \emph{full} pair conservation estimation (using the whole expression in \autoref{eq:pw_pc}) of $v_r$ give equivalent results.
    \item The direct estimation of $v_r$ is more affected by finite $N$ noise than the pair conservation estimation. This is a direct result of the larger variance of $v_r$ with respect to the 2PCF.
    \item The $v_r$ statistic is more affected by finite boxsize effects (small L) than the 2PCF. Thus, we can observe deviations from self similarity at smaller scales in the former statistic than in the latter.
\end{itemize}

\subsection{Resolution as a function of time}

\begin{figure*}
    \begin{subfigure}{\linewidth}
        \includegraphics[width=\linewidth]{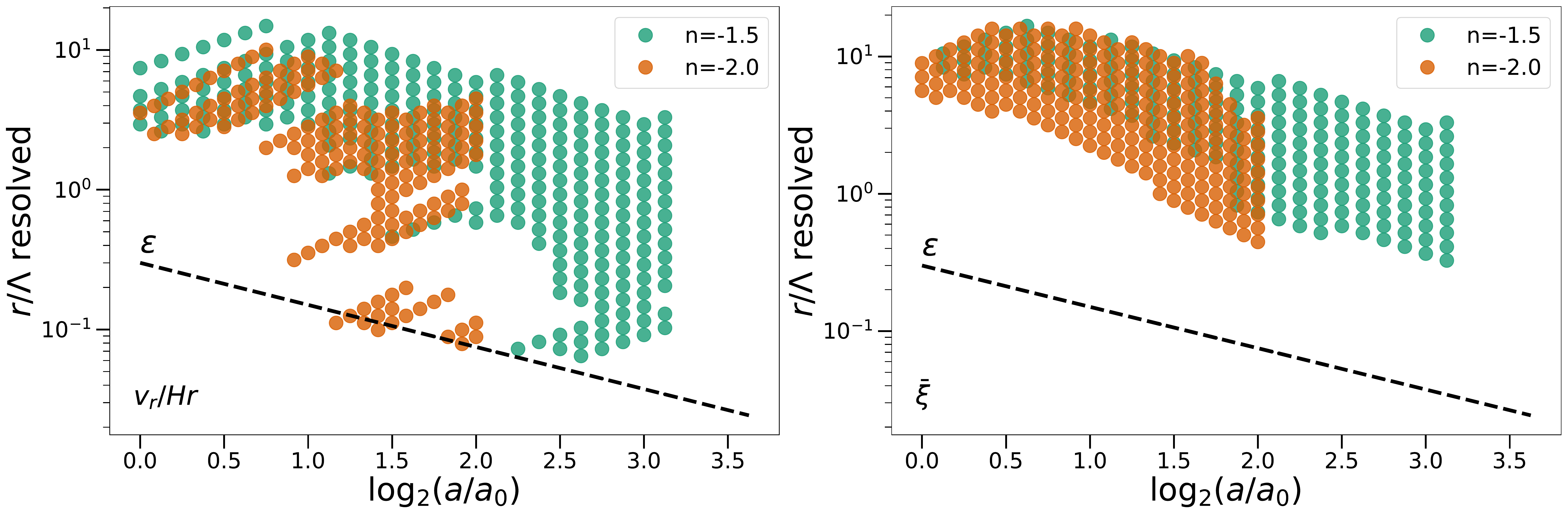}
    \end{subfigure}  
    \begin{subfigure}{\linewidth}
        \includegraphics[width=\linewidth]{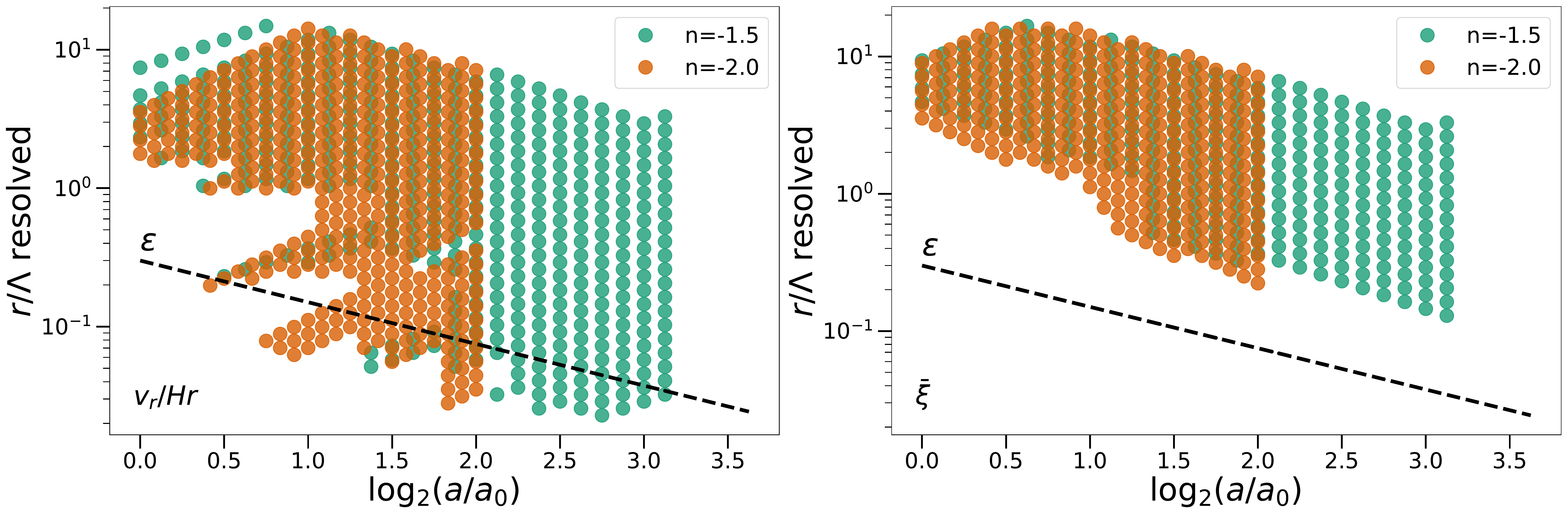}
    \end{subfigure}  
    \caption{Resolved scales (in units of the initial grid-size, $\Lambda$) at $1\%$ (upper row) and $5\%$ (lower row) precision as a function of $\log_2(a/a_0)$. We show results for the 
    spectral indices $n=-1.5$ and $n=-2.0$ (in green and orange, respectively) using the simulations with $N=4096^3$ for the former and the average of four $N=1024^3$ simulations for the latter. The left panels show the results for $v_r/Hr$ (direct estimation), while the right shows the results for the \emph{cumulative} 2PCF. The black dashed line shows the evolution of the softening-length $\epsilon$ in units of $\Lambda$ (which is the same in all simulations).}
    \label{fig:Res_lim}
\end{figure*}

Applying the analysis detailed above to all bins, we can deduce the comoving scales that are resolved (i.e. self-similar) at each given time, for each of the statistics and estimators we have calculated. 

\autoref{fig:Res_lim} shows the comoving separation, in units of the grid spacing, of the resolved bins at the $1\%$  (upper two panels) and $5\%$ (lower two panels) precision levels, i.e. of the bins found to be converged according to the criteria described in \autoref{sec:converge_method} for $p=0.01$ and $p=0.05$. The points in the left panels are for the mean pairwise velocity direct estimate using the $N=4096^3$ simulation for $n=-1.5$ and the average over the four $N=1024^3$ simulations for $n=-2$. The right panels show the \emph{cumulative} 2PCF using the same simulations.

The resolution ranges for $\bar{\xi}$ (in the right panels) can be taken essentially to be those for the mean pairwise velocity estimated from pair conservation and imposing the additional constraint that $\bar{\xi}$ is resolved, i.e. $\dot{\bar{\xi}}=0$, because $\xi(r)$ is always resolved starting from a significantly smaller scale than for $\bar{\xi}$ as can be seen in the right panels of \autoref{fig:SS_n15} and \autoref{fig:SS_n20}. This is just a simple consequence of the fact that $\bar{\xi}$, by definition, is sensitive (at any given precision level) to  $\xi(r)$ over a range of scale below $r$. It will only therefore be resolved starting from a lower cut-off, below which $\xi(r)$ is resolved over some significant range.

Comparing the upper panels, we see that the scales at which $v_r$ is resolved from direct estimation (left panel) and would be from a \emph{reduced} pair conservation estimation (imposing $\dot{\bar{\xi}}=0$) are very similar. A relaxation of the self-similarity constraint in the \emph{cumulative} 2PCF would extend only very modestly the resolved regions, and only at very late times, at least for the case of convergence at the $1\%$ level. There are some additional bins that meet the convergence criterion for the direct estimator, but most of them are not contiguous with the main converged region and thus do not actually extend the lower limit to resolution (i.e. the scale below which convergence is affected by the unphysical UV scales).

In contrast, at $5\%$ precision (lower panels), there is a very marked difference between the two plots: as anticipated from our more qualitative analysis above, we see that the resolution of the pairwise velocity now extends down to scales of order the softening length (indicated by the dashed line in each plot). As we will discuss further below, the apparent explanation for this is that the behaviour of the pairwise velocity at these small scales  --- corresponding to stable clustering --- remains the same whether the spatial clustering is resolved or not.

\subsection{Resolution limits extrapolated to LCDM}

LCDM models are not scale-free:  the linear PS is not a power-law, and there are deviations from EdS power-law scale factor. Nevertheless, the latter deviations are only at very low redshift and 
the PS, in the range of scales relevant to large scale structure formation in cosmology, can be well approximated as a slowly varying power-law: its logarithmic slope varies roughly between $n=-2.5$ and $n=-1.5$ over two decades in scale. From \autoref{fig:Res_lim} we see that the behaviour of the lower cut-off to resolution is quite weakly dependent on $n$ when plotted as a function of $a/a_0$. Thus, we can confidently bracket the lower resolution limits (due to the $UV$ cut-offs, $\Lambda$ and $\epsilon$) using the scale-free results.  

\begin{figure}
    \centering
    \includegraphics[width=\linewidth]{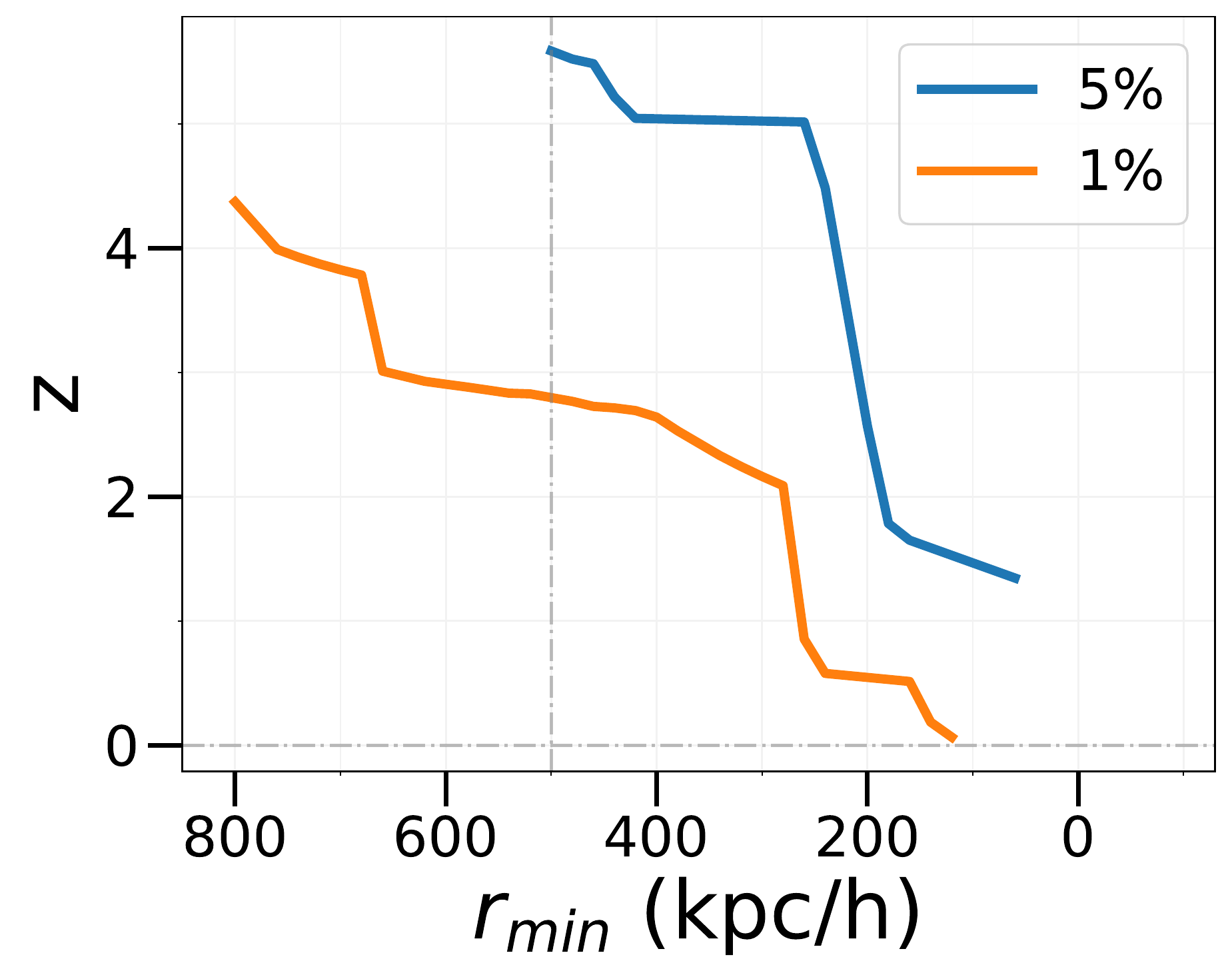}
    \caption{Minimum comoving scale $r_{\rm min}$ at which $v_r/Hr$ is resolved as a function of redshift, estimated for a standard LCDM cosmology (``Planck 2013'', \citet{planck_2013}) in an $N$-body simulation with a mean-interparticle spacing of $0.5$h$^{-1}$ Mpc (indicated by dashed vertical line). The orange (blue) line corresponds to the $1\%$ ($5\%$) precision limits, calculated using data from direct estimation as displayed in the left panels of \autoref{fig:Res_lim}, for the $N=4096^3$ with $n=-1.5$ simulation.}
    \label{fig:LCDM_PW}
\end{figure}

As discussed in our previous analyses (a summary of the relevant information can be found in appendix~\ref{app:copyPS}), for a given physical grid spacing of a LCDM simulation, one can infer $a_0$ and then obtain a conversion between redshift $z$ and the variable $\log_2(a/a_0)$, which allows an approximate ``mapping'' of the scale-free results to the LCDM simulation. Taking the tighter bounds obtained for $n=-1.5$, \autoref{fig:LCDM_PW} shows an example of conservative resolution for a simulation with $\Lambda=0.5h^{-1}\text{Mpc}$. Results are given for a $1\%$ (orange) and $5\%$ (blue) precision in the direct estimation of the pairwise velocity, as plotted in the left panels of \autoref{fig:Res_lim}. Note that the larger missing scales at $5\%$ simply show that $v_r/Hr$ is converged at much earlier redshifts.

\subsection{Converged mean pairwise velocities and stable clustering}

Having focused on identifying the resolved scales, it is also interesting to look at what can be inferred about the behaviour of the studied statistics, and in particular about their behaviour at asymptotically small scales, where the convergence or deviation from stable clustering is of particular interest. 

\begin{figure}
    \centering
    \includegraphics[width=\linewidth]{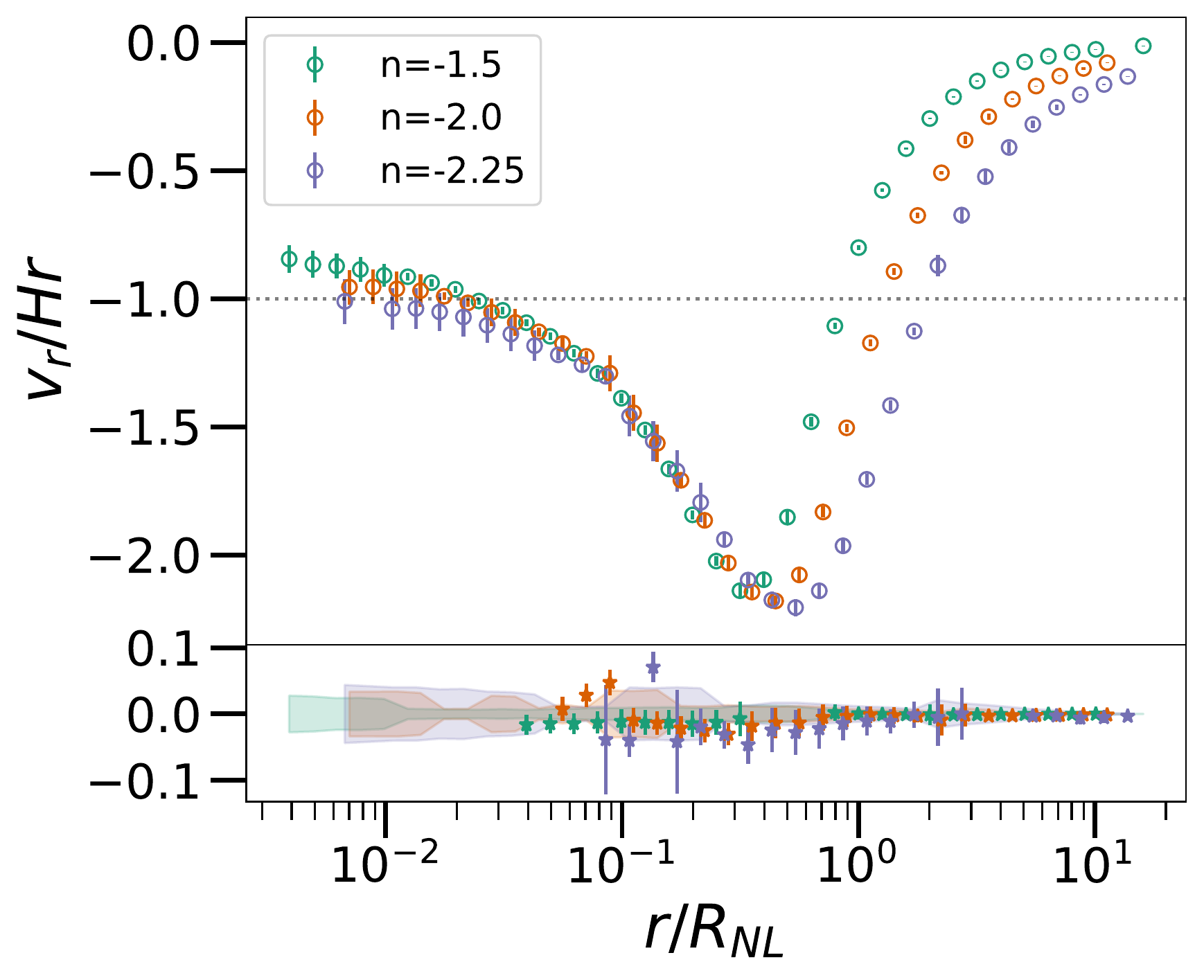}
    \caption{Estimated converged $v_r/Hr$ as a function of rescaled separation, for the three different indicated exponents. In the main plot the same data has been used for the left panels in \autoref{fig:Res_lim} i.e. using direct estimation. The converged values are obtained using the $1\%$ and $5\%$ precision criteria, with the error bars estimated as described in the text. The sub-plot shows these same errors as shaded regions; it also shows (star symbols) the relative difference with the converged values obtained using pair counting estimation.}
    \label{fig:pw_conv}
\end{figure}

We show in \autoref{fig:pw_conv} the converged values of the normalized pairwise velocity for the three simulated spatial indices. These values correspond to the same analysis used to obtain the left panels in \autoref{fig:Res_lim}, but while these show the resolved regions, we now plot the corresponding converged values in each rescaled bin determined by this analysis (i.e. the mean values $X_{\rm conv}$ from the discussion in \autoref{sec:converge_method}). The points plotted are a combination of the values for the bins converged at the $1\%$ level and at the $5\%$ level: we plot $X_{\rm conv}$ for all bins converging at $p=0.01$, and then also for the bins which do not converge at $p=0.01$ but do at $p=0.05$. We add an indicative estimate of the error on $X_{\rm conv}$ which takes into account the expectation that it will decrease as the size of the converged window increases:
\begin{equation}
    \delta=\pm p \sqrt{\frac{w_{\text{min}}}{w}}
\end{equation}
where $w$ is the size (in consecutive snapshots) of the converged window (used to calculate $X_{\text{conv}}$) and $w_{\text{min}}$ the smallest window for which \autoref{eq:X_conv} is satisfied (we have taken here $w_{\text{min}}=3$). Error bars for the $1\%$ level are smaller than the points, thus where the error bars are visible, the corresponding bins converge only at the $5\%$ level. As could be anticipated,  we see that both the accuracy and range of scale measured increases as $n$ does.  

We see in this plot that, while there is a clear $n$-dependence in the shape of the function at larger scales, the behaviour at asymptotically small scales shows a remarkable consistency towards a ``universal" stable clustering (bearing in mind that the error bars are only quite rough estimates of the systematic uncertainties due to finite resolution). Positing this to be the correct physical limit also explains why it can be measured quite well even at scales where the physical behaviour of the clustering is not itself resolved: stable clustering is a robust behaviour that it is not spoiled by the discretization of the density field in an $N$-body simulation.

%% file: Sections/6_Conclusions.tex
The analysis we have reported here is an extension of that in a set of papers \cite{Joyce2021,Leroy2021,Garrison2021,Garrison2021c,Maleubre2022}, which have shown the usefulness of self-similarity and scale-free cosmologies in quantifying resolution of cosmological $N$-body simulations. Our focus here
has been on the radial component of the pairwise velocity in the full matter field. We have also extended, as a complement and for comparison, the analysis of the 2PCF of the matter field (previously studied in \cite{Joyce2021, Garrison2021}). Compared to these previous studies which used a single power law ($n=-2.0$) and simulations of a single size ($N=1024^3$), as in \cite{Maleubre2022} we have considered a set of both different power laws and different box sizes. Unsurprisingly, we have found that the same methods indeed allow us to quantify the evolution of resolution at small scales of the mean pairwise velocity, and further confirm the high levels of accuracy attained by the \Abacus code, also in its determination of correlations in the velocity field. 

Our exploitation here of simulations of different sizes, of several IC realizations, and of scale-free models with different exponents has allowed us not only to improve some of the results in previous work but has also been essential to allow us to extend the method to a velocity statistic.
This is the case because it is crucial for an accurate determination of the precision of convergence 
to be able to separate very clearly the effects of discretization at small scale from both the 
noise and systematic effects at large scales due to the finite box size. For the pairwise velocity 
statistics, which are more sensitive than the 2PCF to these effects, the comparison of different 
(and larger) box sizes and different exponents turns out to be essential to disentangle clearly the 
different effects. We have also exploited the two different estimators of the velocity statistic --- directly from the particles' velocities in the simulation or indirectly by pair-counting  --- to identify noise due to finite size effects.
The comparison of different exponents has allowed us also to see how the range of converged scales markedly degrades due to finite size effects as $n$ decreases, and in practice our $n=-2.25$ simulations are not useful for placing precision limits at the $1\%$ level. Further, we argue that our results for the evolution of small scale resolution can be extrapolated to LCDM type models, as they are, when suitably expressed, very weakly dependent on scale-free index $n$ (which values have been chosen to probe the relevant 
range in LCDM). The same is not true of box size effects, which are strongly $n$ dependent, and indeed we do not attempt to make an extrapolation for these.

We have found that we can determine the evolution of
lower cut-off to resolution at the $1\%$ level for the radial pairwise velocity. In addition, we show that it is approximately equal to the corresponding 
cut-off for the \emph{cumulative} 2PCF, which converges at the same precision level varying from a few times
the grid spacing at early times to slightly below this scale at late times.
This is a few times larger than the scale at which the 2PCF itself attains the same precision  
\cite{Joyce2021, Garrison2021}. This reflects the coupling of the velocity correlation
at a given scale to the clustering at smaller scales (as expressed through the integral
$\bar{\xi}$ in the self-similar limit). 

On the other hand, at $5\%$ precision we have 
obtained resolution extending down to scales of order the softening length, $\epsilon$,
where even the 2PCF is far from its converged value \cite{Joyce2021, Garrison2021}. 
In the corresponding range of scale $v_r/Hr \approx -1$, i.e. the result 
is consistent with the so-called stable clustering hypothesis 
in which non-linear structures become stationary in physical coordinates \cite{Peebles1974}. 
The conclusion that clustering may indeed tend to this behaviour at asymptotically 
small scales is consistent with an early analysis (with much smaller simulations,
$N\sim 10^6$) of the question using pairwise velocities by \cite{Jain1997} 
(estimated by pair-counting), and also with results for the shape of the 
power spectrum at large $k$ reported in \cite{Maleubre2022}. In this hypothesis, 
the fact that resolution extends to such small scales for  $v_r/Hr$ is simply due 
to the fact that the stable behaviour is not spoiled by the discretization of the 
matter field, and persists even if the clustering is very different to that 
in the continuum model.

With respect to the preparation of theoretical predictions for forthcoming surveys, and specifically for
redshift space distortions, our analysis of the pairwise velocity gives only an indication of the resolution 
limits at small scales in $N$-body simulations. It would be straightforward to extend our analysis to
additional statistics used in this context, e.g. PDFs of the pairwise velocity and their moments (see references in introduction).
Further, to attain a quantification of bounds for the typically cited target $1\%$ level would require slightly more data sets than what we have used here ---- either slightly larger simulations, or a couple of realizations of the same size as our largest simulations here. 

We conclude with some comments on other possible further developments of this work. 
Our analysis of the mean pairwise velocities in the dark matter field (cf. \autoref{fig:pw_conv}) shows an apparently 
universal shape below the scale of maximal infall, and going asymptotically to stable clustering.
It would be interesting to compare these results with those in LCDM, making use of the resolution 
limits we have determined here, to assess whether we indeed find the same behaviour.
To establish the evidence for stable clustering at asymptotically small scales, a fuller comparative
joint analysis of the 2PCF, PS, and pairwise velocity itself should be performed.

%% file: Sections/Appendix.tex
Following the steps of our previous studies using scale-free simulations, we have characterized how resolution depends on time in terms of a scale factor relative to $a_0$, corresponding to a characteristic time at which non-linear structures start to develop in a simulation. For any cosmology (e.g., both EdS and LCDM-like), it can be defined as given by \autoref{eq:def_a0}, which simply relates it to the value of the variance at a given scale (in this case the mean particle separation $\Lambda$ in a simulation). Thus, the mapping between EdS and LCDM time evolution is just a function of the mean interparticle spacing 
$\Lambda$ and the linear power spectrum of the model, as these allow the determination of $a_0$.

\begin{figure}
 \centering
 \includegraphics[width=1.\columnwidth]{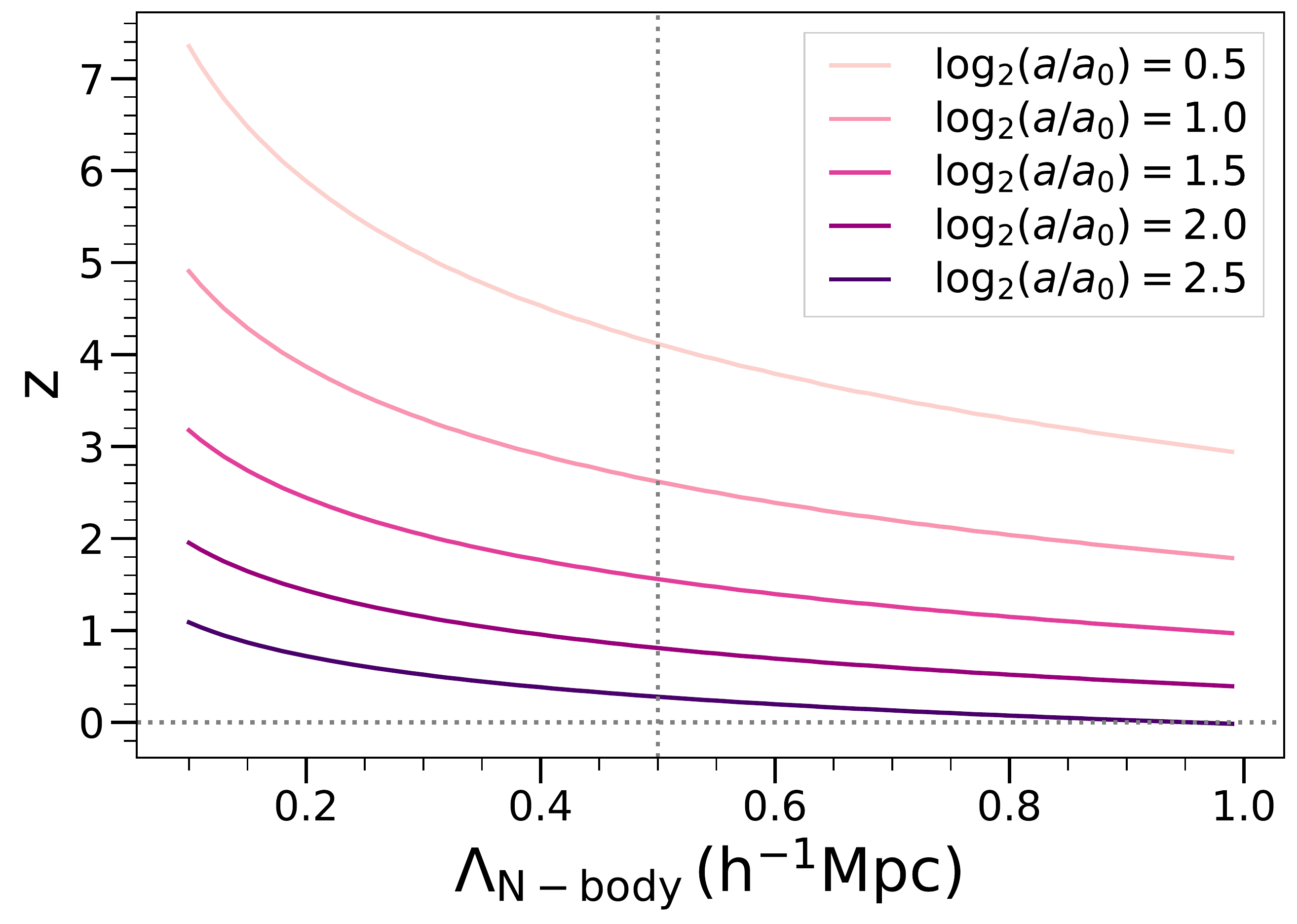}
 \caption{Redshift $z$ corresponding to different fixed values of $\log_2(a/a_0)$ as a function of mean interparticle spacing $\Lambda$, using a standard LCDM cosmology (``Planck 2013'', \protect\cite{planck_2013}). As discussed in the text, combining this plot with the curves from \autoref{fig:Res_lim} we can infer a conservative bound on attainable precision as a function of redshift in an LCDM simulation.}
 \label{fig:zeff}
\end{figure}

\autoref{fig:zeff} illustrates how the parameter 
$\log_2(a/a_0)$ maps to the redshift in a simulation of a standard LCDM model (``Planck 2013'', \cite{planck_2013}). This means that, given a simulation of a determined grid spacing $\Lambda$, one can always find a one-to-one relation between the desired evolved redshift of the LCDM and our time variable $\log_{2}(a/a_0)$. As discussed in \citet{Joyce2021}, non-EdS expansion at low redshift introduces the possibility of mapping the time rather than the scale-factor, but the difference in the effective $\log_2(a/a_0)$ is in practice very small, and we will neglect it here.

This mapping allow us to extrapolate the minimum scales from \autoref{fig:Res_lim}, found in EdS cosmologies, into the scales in \autoref{fig:LCDM_PW}, for a LCDM simulation. 

